\documentclass[noshowpacs,prl,preprint,preprintnumbers,superscriptaddress,amsmath,amssymb,floatfix]{revtex4}
\usepackage[caption=false]{subfig}
\usepackage{graphicx}
\usepackage{placeins}
\usepackage{mathtools}
\usepackage{amsthm}
\usepackage{romannum}
\usepackage{color,soul}
\usepackage{xkeyval,xcolor}

\usepackage{algcompatible}
\usepackage{newfloat}
\DeclareFloatingEnvironment[
fileext=loa,
listname=List of Algorithms,
name=ALGORITHM,
placement=tbhp,
]{algorithm}

\algtext*{EndWhile}
\algtext*{EndIf}

\begin{document}
	
	\title{Percolation framework reveals limits of privacy in Conspiracy, Dark Web, and Blockchain networks}
	\author{Louis M. Shekhtman}
	\affiliation{Network Science Institute, Northeastern University, Boston, USA}	
	\author{Alon Sela}
	\affiliation{Department of Industrial Engineering, Ariel University, Ariel, Israel}
	\author{Shlomo Havlin}
	\affiliation{Department of Physics, Bar-Ilan University, Ramat Gan, Israel}
	\date{\today}
	\begin{abstract}
	We consider the privacy of interactions between individuals in a network. For many networks, while nodes are anonymous to outside observers, the existence of a link between individuals implies the possibility of one node revealing identifying information about its neighbor. Moreover, while the identities of the accounts are likely hidden to an observer, the network of interaction between two anonymous accounts is often available. For example, in blockchain cryptocurrencies, transactions between two anonymous accounts are published openly. Here we consider what happens if one (or more) parties in such a network are deanonymized by an outside identity. These compromised individuals could leak information about others with whom they interacted, which could then cascade to more and more nodes' information being revealed. We use a percolation framework to analyze the scenario outlined above and show for different likelihoods of individuals possessing information on their counter-parties, the fraction of accounts that can be identified and the idealized minimum number of steps from a deanonymized node to an anonymous node (a measure of the effort required to deanonymize that individual). We further develop a greedy algorithm to estimate the \emph{actual} number of steps that will be needed to identify a particular node based on the noisy information available to the attacker. We apply our framework to three real-world networks: (1) a blockchain transaction network, (2) a network of interactions on the dark web, and (3) a political conspiracy network. We find that in all three networks, beginning from one compromised individual, it is possible to deanonymize a significant fraction of the network ($>50$\%) within less than 5 steps. Overall these results provide guidelines for investigators seeking to identify actors in anonymous networks, as well as for users seeking to maintain their privacy. 
	\end{abstract}

\maketitle

The rise of privacy concerns in modern society has led to an increased interest in possible methods for individuals to interact while preserving their anonymity \cite{gross2005information,garcia2017leaking,bagrow2019information,reid2013analysis,lazer2009computational,song2010limits, onnela2007structure}. These individual interactions form a network between different parties and in many cases, while the anonymity and privacy of an individual node is preserved, the global structure of the network may be visible to others \cite{zheleva2009join}. Often times the very act of interacting via a network link will involve the two individuals at the ends of the link exchanging  information that allows them to identify one another \cite{xu2004fighting}. Thus, if one individual is identified, it may (with some probability) be possible to obtain information on some of her connections in the network \cite{barucca2018tackling}. 

For example, if a physical item is purchased via an anonymous transaction network e.g., Blockchain cryptocurrencies \cite{pappalardo2018blockchain}, then the buyer must provide a shipping address to receive the item and thus reveals some aspect of their location. Likewise, if two individuals communicate via phone, then it is likely that they possess some knowledge of the person on the other end of the call. Finally, in the case of online interactions, if  user A's computer is hacked then A's identity will become known to the hacker. The hacker could then search for additional information on A's computer e.g., email correspondence or private messages through online forums, to learn the identities of other individuals who interacted with A. After doing so, the hacker could then attempt to hack into these individuals computers (e.g., via a Trojan horse email) and with some probability, successfully traverse the network. The very fact that A was hacked could also be useful in hacking the neighbors of A as individuals are more likely to trust emails (that might contain a hidden Trojan) from known sources \cite{moody2017phish}. Similarly, in the context of criminal, terrorist or conspiracy networks, if a set of individuals interact via anonymous communications if one member of the group is identified, they could potentially be monitored or interrogated leading to the revelation of information on other individuals, who could then also be monitored  to identify other members of the network, and so on. Specific examples would include users of burner phones who do not provide their names, but whose calls can be tracked; anonymous email accounts where a name is not provided, but a user's messages may be saved by the email provider; Telegram Messenger, where anonymous users may interact in public groups; and other criminal or conspiracy networks.

Here we show that the question of anonymity between network actors, and the corresponding ability of a party seeking to deanonymize the individuals based on information from their neighbors, can be solved using tools and methods of percolation from statistical physics \cite{newman2010networks,cohen2010complex,castellano2009statistical,barabasi2016network}. Furthermore, we demonstrate that classical quantities from statistical physics have important meanings and provide crucial information on the scope to which anonymity can be maintained among individuals in real hidden networks. 

We focus on three example networks: (1) transactions through blockchain-based cryptocurrencies \cite{chen2018understanding}, (2) interactions related to illegal activities via the dark web \cite{da2020assessing}, and (3) a political conspiracy network \cite{ribeiro2018dynamical}. In all three of these networks, the question of anonymity is very important- as cryptocurrencies and the dark-web are often used by criminal organizations, \cite{ron2013quantitative,ober2013structure,meiklejohn2013fistful,de2017modeling} whereas conspirators depend on not being uncovered in order to avoid criticism and possible criminal charges.

\begin{figure*}
	\centering
	\includegraphics[width=0.95\linewidth]{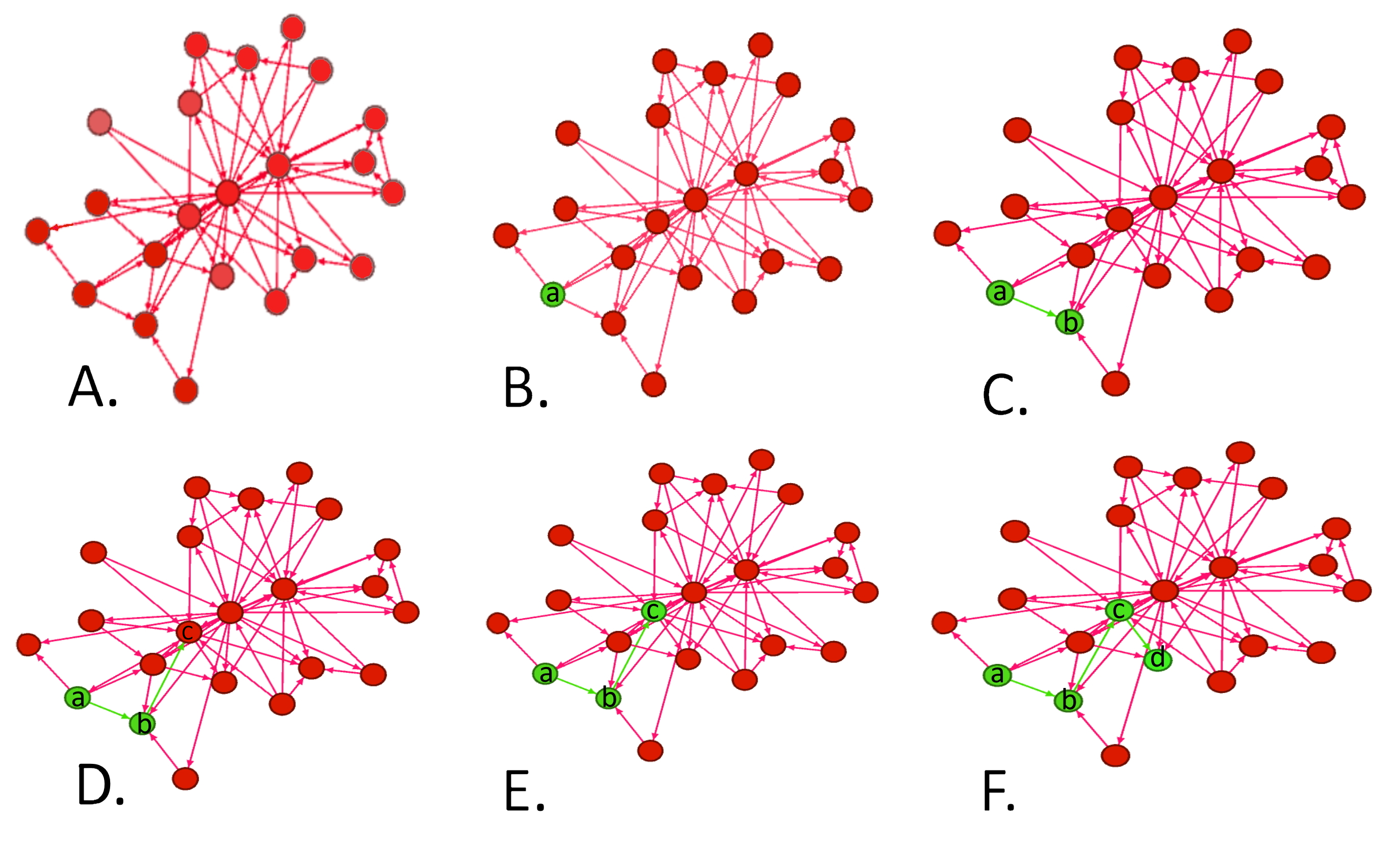}  	
	
	\caption{\textbf{Giant Component and deanonymization.}  The initial network (A) where no users are known. In (B) after investigators successfully identify node \emph{a}, this node is marked green meaning it is known and investigators can then use information from node \emph{a} to traverse the network. Initially all links are not known thus marked red. Green links represent links that have been  de-anonymized. In (C) the identity of node \emph{b} is identified through information obtained via a link to node \emph{a}. Then, in (D), information from node \emph{b} provides a link to node \emph{c}, who is also identified in subplot (E). Finally, node \emph{c} provides information resulting in the identification of node \emph{d} in subplot (F).} 
	
	
	\label{fig:network-traverse}
\end{figure*}

While some approaches have considered deanonymizing the individuals behind nodes in a network \cite{ron2013quantitative}, these have typically related to specific encrypted protocols and users have found ways to overcome these issues. In contrast, our approach is fundamental to the nature of privacy when interacting in a network i.e., when interacting with another party, a user often must reveal some identifying information. Using our general framework demonstrated in Fig~\ref{fig:network-traverse} and based on the topologies of three real hidden networks, we are able to quantify the extent to which this information can be exploited and thus reveal how individuals can be unwittingly identified by their neighbors who failed to remain anonymous.

\section{Results}

As explained above, we seek to understand how investigators could leverage information exchanged between interacting parties in order to identify them. Nonetheless, in some cases, an individual may not have any identifiable information about the party at the other end of the transaction. In this sense, one could consider such a link `failed' in the sense that no deanonymization can be carried out via that link. Thus, the fraction of links exchanging identifying information can be mapped to the link-occupation probability $p$ from percolation, such that $p$ fraction of links exchanged identifying information. Traditionally, in percolation frameworks the key quantity of interest is then the fractional size of the giant connected component $S$. In the context of anonymity, if a single node in the giant component is deanonymized, then the entire giant component could be discovered (within some time) via individuals revealing information about their neighbors. 

Similarly, for the other components, we can state that each component is a set of nodes where if any one of these is deanonymized, then the rest of the set could also be identified. Therefore the total number of components, $n_{comp}$, represents the minimal number of source nodes (in distinct components) that need to be identified independently in order to deanonymize the entire network. Similarly if one has a set of source nodes $n_{1,2,...}$ each in a different component, then the total number of nodes that can be identified is the sum of the sizes of all of those separate components $T_S=|S_1|+|S_2|+...+|S_n|$, where $T_S$ is the total number of nodes that can be identified (see SI- Distribution of Component Sizes). 

Also of significant importance for privacy is the actual number of interrogations that an investigator must carry out in order to deanonymize a specific target node, given that the investigators know the identity of some given source nodes. For example, perhaps investigators identify one dark web user posting large amounts of child pornography or a specific account participating in suspicious transactions on the Blockchain. The investigators could begin by seeking information from their source individual on her neighbors and then move on to the neighbors' neighbors and so on, until reaching the desired node of interest. This minimum number of individuals that must be identified to reach the desired node is the shortest path length, $\ell$, from the source node/s to the target node and it represents a measure of effort the investigators must expend since each individual that must be identified along the path will require dedication of resources.. Lastly, we propose a new measure related to the fact that while the investigators presumably know that some individuals will not have identifying information on all of their neighbors, they do not know in advance which links will be helpful. Therefore, we propose a greedy algorithm, described later, where the investigators first interrogate the nodes along the initial shortest path from their source node/s to the target and then update to a new path if the investigation reaches a dead end; i.e. a link where the node on the other end cannot be identified. We define the number of steps along the path using our greedy algorithm as $\ell_{actual}$ as it approximates what could be taken as a possible `actual' number of steps needed to reach a specific node given that investigators do not know which links are useful. These and other relevant measures from percolation theory are described in Table~\ref{table:definitions}.


\begin{table}
	\centering
	\begin{tabular}{||c | c | c||} 
		\hline
		Parameter & Percolation Theory Definition & Interpretation \\ [0.5ex] 
		\hline\hline
		$p;p=1-q$ & Fraction of occupied links & Fraction of links where nodes can identify one another  \\ 
		\hline
		$S$ & Giant connected component (GCC) & Largest group of mutually vulnerable nodes  \\ 
		\hline
		$SG$ & Second largest connected component & Second largest group of mutually vulnerable nodes  \\ 
		\hline
		$n_{comp}$ & Number of Components & Min. num. of source nodes to identify whole network \\
		\hline
		$\ell$ & Shortest Path Length & Optimal number of interrogations \\
		\hline
		$\ell_{actual}$ & Greedy Algorithm Path Length & Realistic number of interrogations \\ [1ex] 
		\hline
	\end{tabular}
	\caption{Mapping to percolation. We demonstrate and provide brief explanations on the various measures used and how they relate to the traditional measures from percolation theory.}
	\label{table:definitions}
\end{table}

We apply the framework described above to three publicly available hidden networks: (i)  the flow of funds within the Ethereum blockchain-based cryptocurrency \cite{chen2018understanding}, which is the second largest cryptocurrency by market cap; (ii) A forum of users participating in sharing of child pornography on the dark web \cite{da2020assessing}; and (iii) A network of political conspirators in Brazil \cite{ribeiro2018dynamical}. Summary statistics on these datasets are available in Table ~\ref{tab:data}.

\begin{table}
 \begin{tabular}{|c | c|  c| c|}
 \hline
  Metric/Network& Blockchain & Dark Web & Conspiracy \\ [0.5ex] 
 \hline\hline
 N & 2,291,941 & 10,407 & 404 \\ 
 \hline
V & 5,262,468 & 820,272 & 3350 \\
 \hline
 $<k>$ & 2.8 & 150 & 9.7 \\
 \hline
 $c$ & 0.21 & 0.83 & 0.85 \\
 \hline
 $<\ell>$ & $4^*$ & 2.15 & 2.98 \\
 \hline
 Density & $10^{-6}$ & 0.0076 & 0.022 \\ [1ex] 
 \hline
\end{tabular}

 \caption{Data used and basic real-world network characteristics. $N$ is the number of nodes, $V$ is the number of links, $<k>$ is the average degree, $c$ is the clustering coefficient, $<\ell>$ is the average shortest path length (which for the Blockchain network could only be estimated due to the network dimensions), and density is the overall density of links defined by the number of links in the network divided by the number of all possible links. \\ $^*$ Median is listed rather than mean due to long tail.}
\label{tab:data}
\end{table}

\begin{figure*}
	\centering

	\subfloat{
		\includegraphics[width=0.315\textwidth, trim=14 0 12 0, clip]{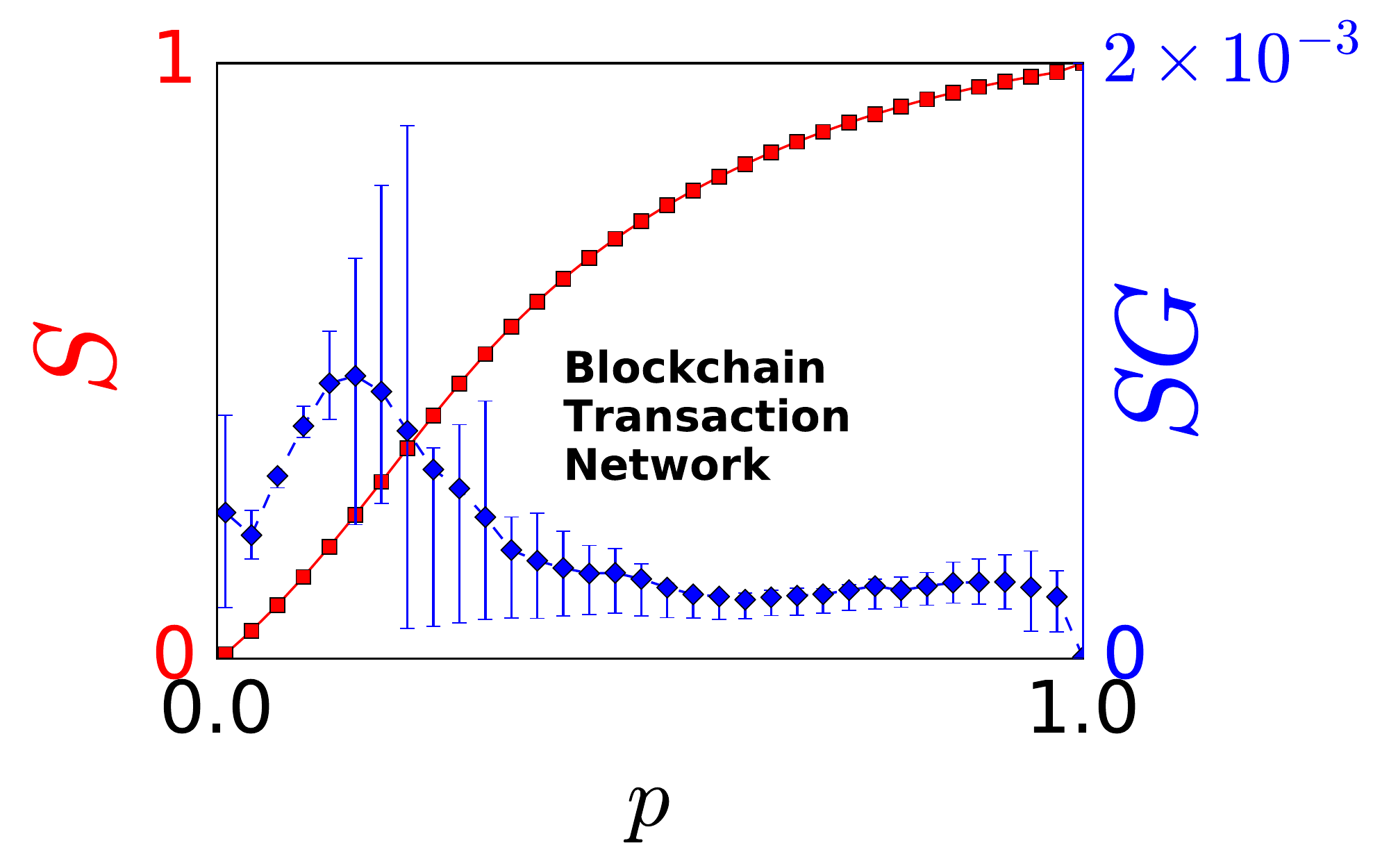}	
		\label{fig:gcc-blockchain}
	}
	\hfill
	\subfloat{
		\includegraphics[width=0.315\textwidth, trim=14 0 12 0, clip]{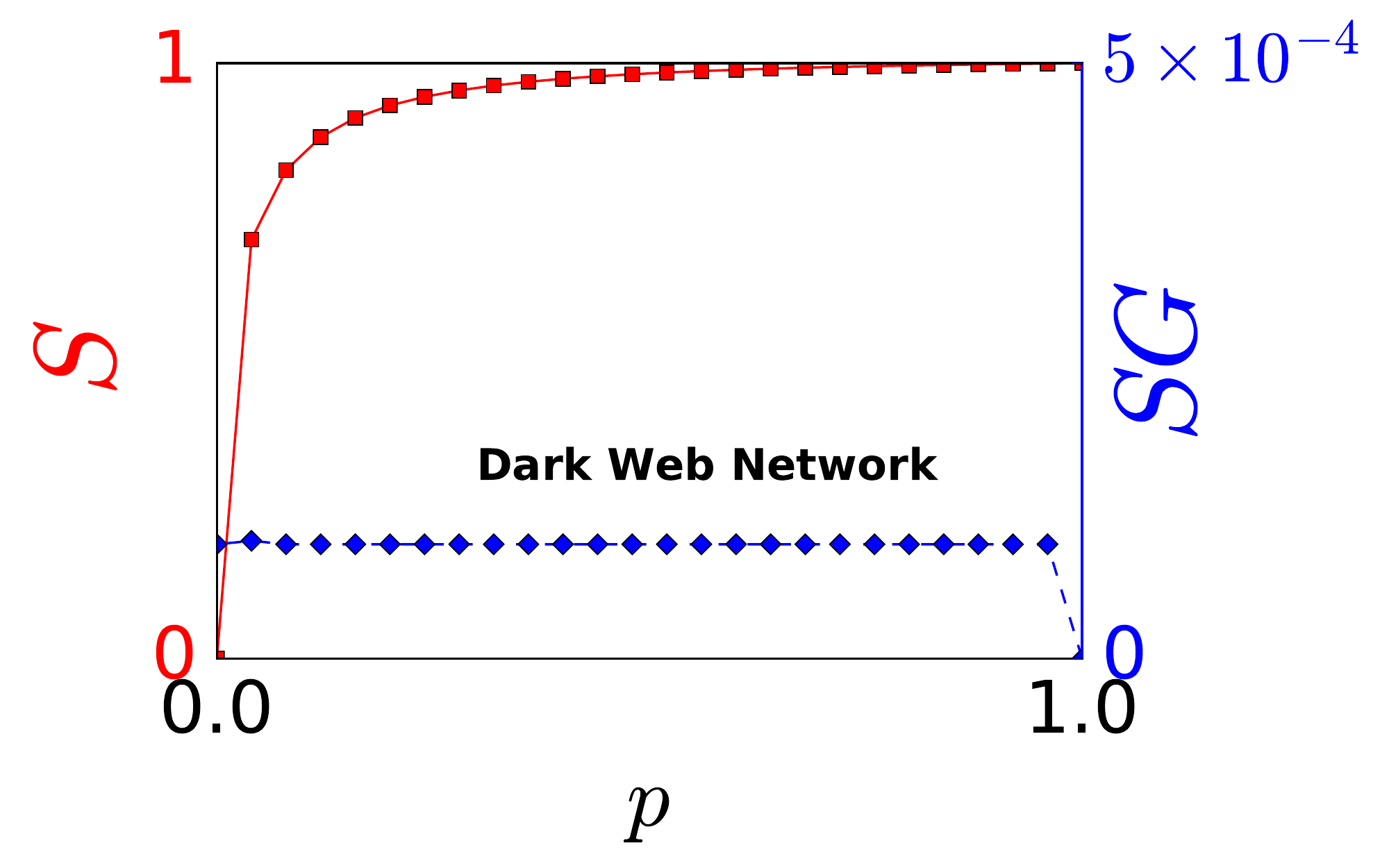}	
		\label{fig:gcc-dacunha}
	}	
	\hfill	
	\subfloat{
		\includegraphics[width=0.315\textwidth, trim=14 0 12 0, clip]{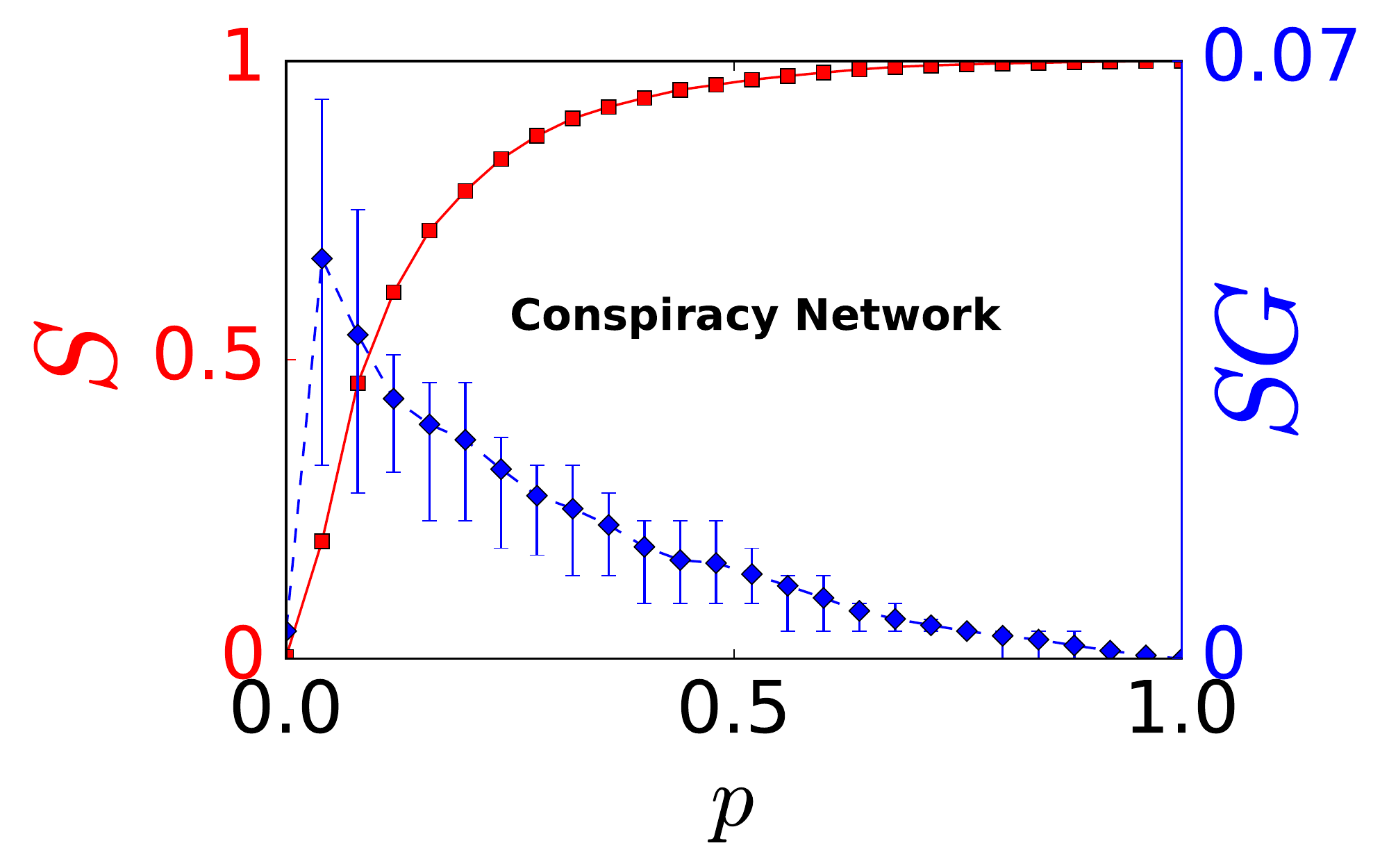}
		\label{fig:gcc-matjaz}
	}

	\caption{\textbf{Largest and Second Largest Component.} We show the fractional size of the largest, $S$, and second largest $SG$, components in each of our three real networks. The points shown are averages of 100 realizations and the error bars represent one standard deviation. For $S$ the error bars are smaller than the size of the markers. We see that the second largest component is generally quite small (right y-axis) compared to the largest component and thus the largest component is most important for deanonymization.}
	\label{fig:gcc-sgcc}
\end{figure*}

Using our percolation approach we analyze the sizes of the largest component in our three real-world networks. This corresponds to the mean number of accounts that can be deanonymized after a single source node is identified. 
In Fig~\ref{fig:gcc-sgcc}, we show the fractional size of the giant connected component, $S$, and the second largest connected component, $SG$, for each of our networks. We see that the second largest component is typically quite small (0.07 at max for the Conspiracy network) in comparison to the largest component whereas $S$ typically constitutes a large fraction of the network. Also we can see that as the probability of deanonymize $p$, increases the giant component increases and the second giant component passes through a peak. The fact that only near $p \to 0$ does $S \to 0$ is typical of networks with long-tail degree distributions, which is true for our 3 networks here (See Fig.~\ref{fig:degree-dist}). 

Overall, these results imply that the key deanonymization efforts should focus on the giant component, and suggest that our approach is highly feasible in terms of  deanonymizing a large fraction of accounts since only a single source node is needed, and then all the nodes in $S$ are potentially identifiable.

\begin{figure*}

	\subfloat{
		\includegraphics[width=0.31\textwidth, trim=7 0 7 0, clip]{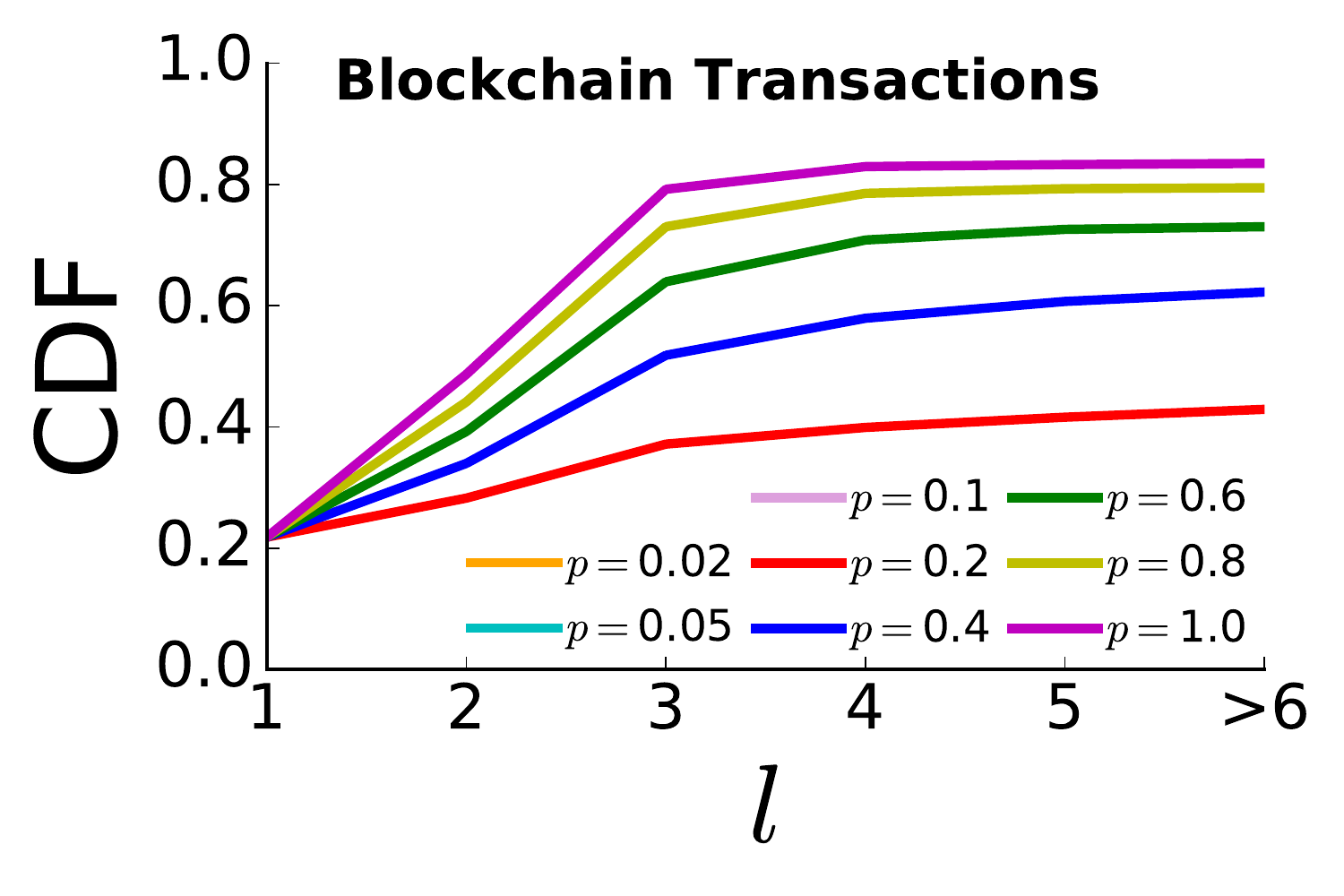}
		\label{fig:paths-blockchain}
	}
	\hfill
	\subfloat{
		\includegraphics[width=0.31\textwidth, trim=7 0 7 0, clip]{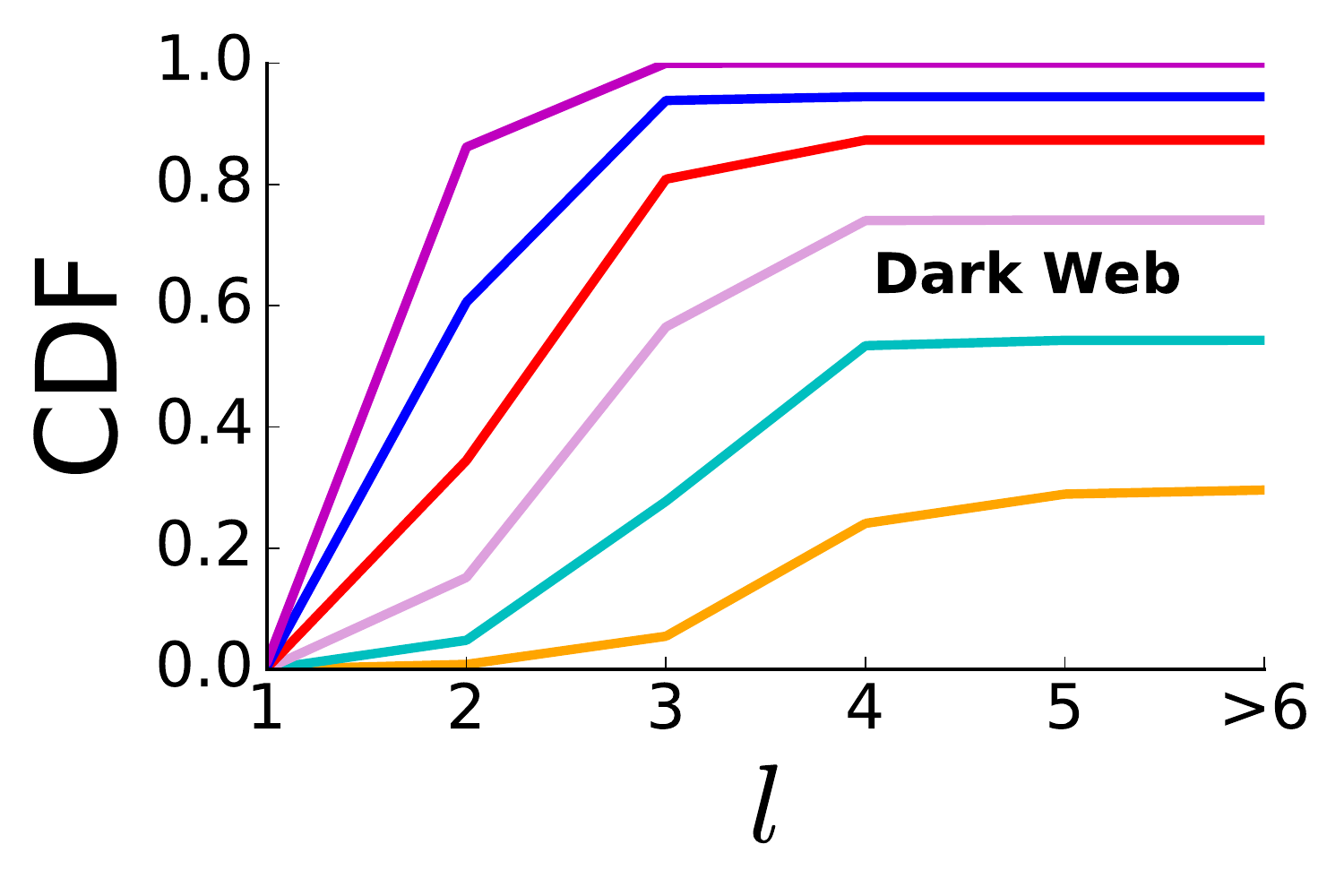}	
		\label{fig:paths-dacunha}
	}	
	\hfill	
	\subfloat{
		\includegraphics[width=0.31\textwidth, trim=7 0 7 0, clip]{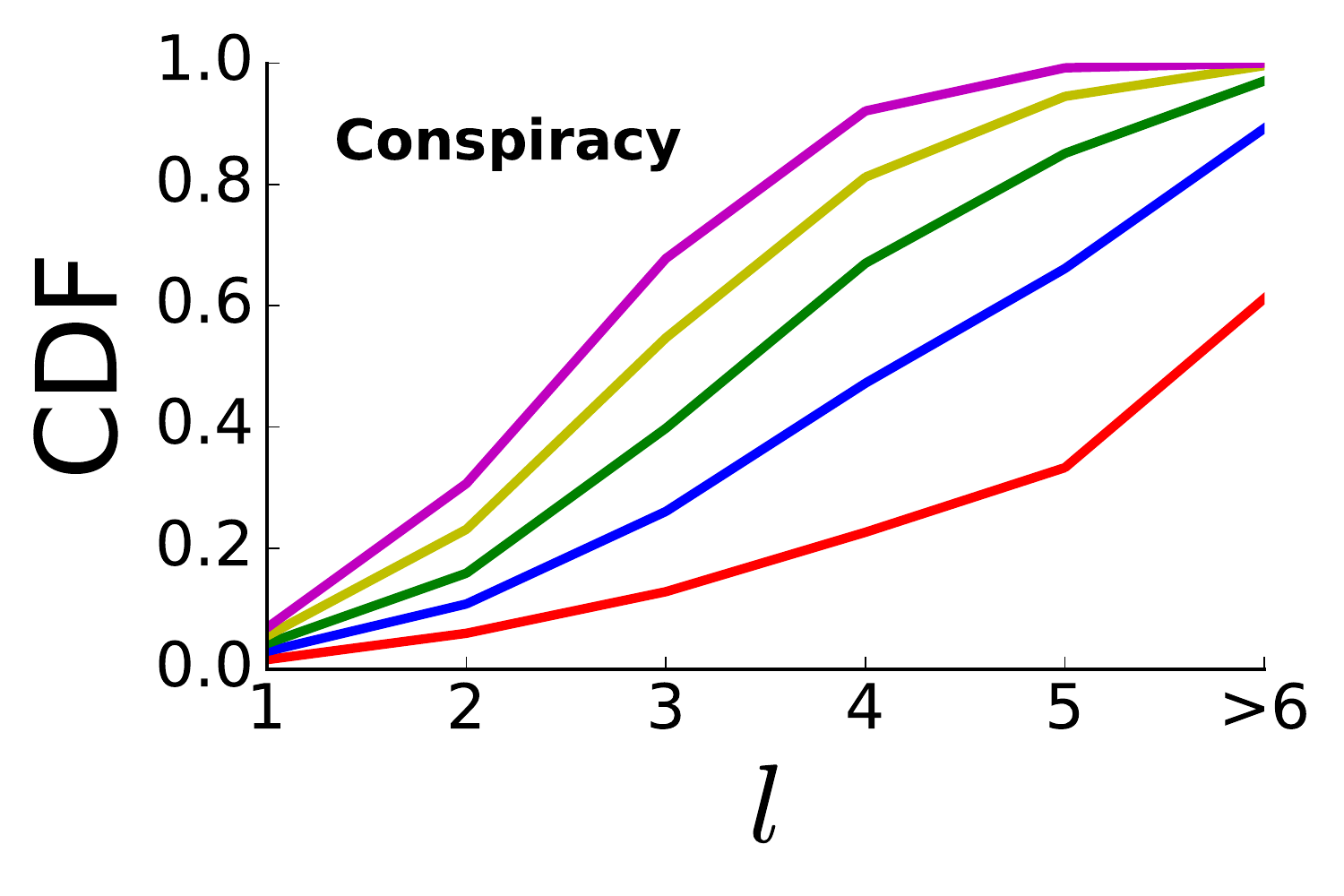}
		\label{fig:paths-matjaz}
	}
	\caption{\textbf{Size within shortest path length (ideal paths).}  We plot the number of nodes that can be reached within $\ell$ steps (shortest path) for each of our networks. For the Blockchain Transaction Network we assume that the set of all exchange nodes are sources, whereas for the Dark Web Network and the Conspiracy Network, we randomly select source nodes.  We see that for all 3 of our networks, the ideal shortest paths are very short, with most reachable nodes, being reached within $\ell\approx3$ or $\approx 4$ steps. Note that values of $p\leq0.1$ are shown only for the Dark Web. }
	\label{fig:spt-cdf}
\end{figure*}

Aside from the {possibility} of identifying individuals, there is also a question of the amount of resources that must be dedicated in order to do so. Therefore an  important aspect of our proposed method is the minimum number of other intermediate individuals that the investigators must identify on the way towards identifying a specific target node. Each identified individual requires considerable  effort such as surveillance,  hacking efforts, interrogations, etc.  This number of intermediate steps will inherently depend on the initial source node where the investigators begin their search and on the topology of the network. In terms of percolation theory quantities, it is related to the shortest path length from the source node to a given target node. To capture an overall picture of the necessary effort, in Fig.~\ref{fig:spt-cdf} we show the cumulative distribution (CDF) of the fraction of nodes that can be found within $\ell$ steps for different values of $p$, and for each of the three networks that are analyzed. We note that for the Dark Web network we explore several very small values of $p$ because interactions on a forum in the Dark Web are less likely to involve individuals having information on one another than in the case of transferring money or cooperating in a conspiracy. The CDF essentially demonstrates for any possible source and target node the range of possible levels of effort needed.

We note that in the case of the Blockchain Transaction Network, rather than considering shortest paths between all pairs of nodes, we begin with a set of known nodes that are so-called `exchanges,' which convert between cryptocurrency and fiat currency. This choice was made because exchanges inherently know the identities of their neighbors due to legal policies they have in place to prevent money laundering \cite{sapovadia2015legal}. Moreover, exchanges are the hubs of the Blockchain Transaction Network \cite{chen2018understanding} and thus serve as most likely initial targets for investigators (for more on this choice, see SI). In contrast, for the Dark Web Network and the Conspiracy Network, there is no initial reason to suspect that any particular node is more likely to serve as an initial source. We see that in all three of our networks the optimal shortest paths are typically within $\ell=3$ or $\ell=4$ steps, meaning that the network can be considered a `small-world' \cite{amaral2000classes} with most nodes being within an identification range of a fairly small number of interrogations. 

\begin{figure*}
	\centering
	\includegraphics[width=0.99\linewidth]{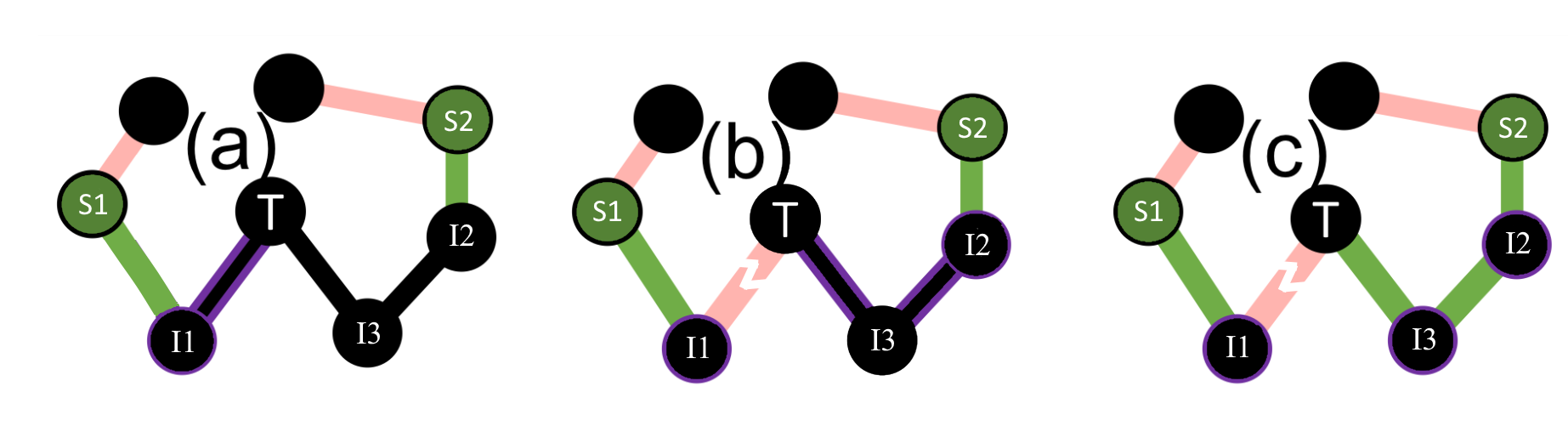}  	
	
	\caption{\textbf{Targeting a specific node.}  Here we demonstrate our greedy algorithm for deanonymizing a specific node $T$, given that we know the identity of two source nodes $S1$ and $S2$. In \textbf{(a)} we find the shortest path from our source nodes which is $S_1\rightarrow I_1\rightarrow T$ (links of the shortest path with unknown identities outlined in purple) and interrogate the first node $I_1$, on this shortest path, for information. \textbf{(b)} After successfully identifying of node $I_1$, we reach a dead end as the node $I_1$ does not provide information on the target node $T$. We then again calculate the shortest path to the target $T$ and find an alternative path from source $S_2\rightarrow I_2\rightarrow I_3 \rightarrow T$ (again with its links to unknown nodes highlighted in purple). \textbf{(c)} We interrogate the first and second nodes on the path; $I2$ and then $I3$ (outlined in purple again), who each successfully provide information, ultimately leading us to the target. Note that while the optimal path would have interrogated only $\ell=2$ nodes, the actual path uses $\ell_{actual}=3$ nodes.}
	\label{fig:node-target}
\end{figure*}

However, the shortest path length is only the \emph{minimal} number of interrogations an investigator would need if they had perfect knowledge in advance about which interactions involved identifiable information. On a practical level, the investigator will be confronted with the `noise' of links that occurred between individuals but did not involve identifying information. Therefore, we propose a naive greedy algorithm (Algorithm~\ref{alg:greedy}) that investigators could use in order to carry out their investigation. Essentially, our greedy algorithm calculates the shortest path between the source node/s and the target. The investigators then interrogate the nodes along the shortest path until they hit a dead end i.e., reach a node that cannot communicate identifying information about the desired neighbor. They then remove that link from the network, calculate the new shortest path to the target node in the modified network, and attempt to traverse the new shortest path. This process can be done iteratively until the target node is reached or until it is known that no other possible paths exist. We note that this algorithm is not necessarily optimal, but on the contrary, provides a simplistic upper bound on the effort needed. 

\begin{algorithm}
	\caption{Greedy Algorithm to Identify Target Node}
	Given graph $G$, set of source nodes $S$, target node $T$.
	\begin{algorithmic}
		\STATE{interrogations=0}
		\WHILE{target node not reached}
		\STATE{Calculate shortest path SPT }
		\FOR{link \ in SPT}
		\IF{link identifiable}
		\STATE{interrogations+=1}
		\ENDIF
		\IF{link unidentifiable}
		\STATE{path=failed}
		\STATE{link\_identifiable[link]=false}
		\ENDIF
		\IF{target node reached}
		\STATE{path=succeeded}
		\STATE{total\_interogations=interrogations}
		\ENDIF
		\ENDFOR
		\ENDWHILE
	\end{algorithmic}
	\label{alg:greedy}
\end{algorithm}

In Algorithm~\ref{alg:greedy} we formally present our greedy algorithm and in Fig.~\ref{fig:node-target} we demonstrate the process of our algorithm visually. If all of the links on the original shortest path are indeed identifiable then our algorithm will lead to the minimal number of interrogations. In contrast, if we hit dead ends along the paths we pursue, then our algorithm will lead to a greater number of interrogations than optimal.

\begin{figure*}
	\centering

	\subfloat{
		\includegraphics[width=0.31\textwidth]{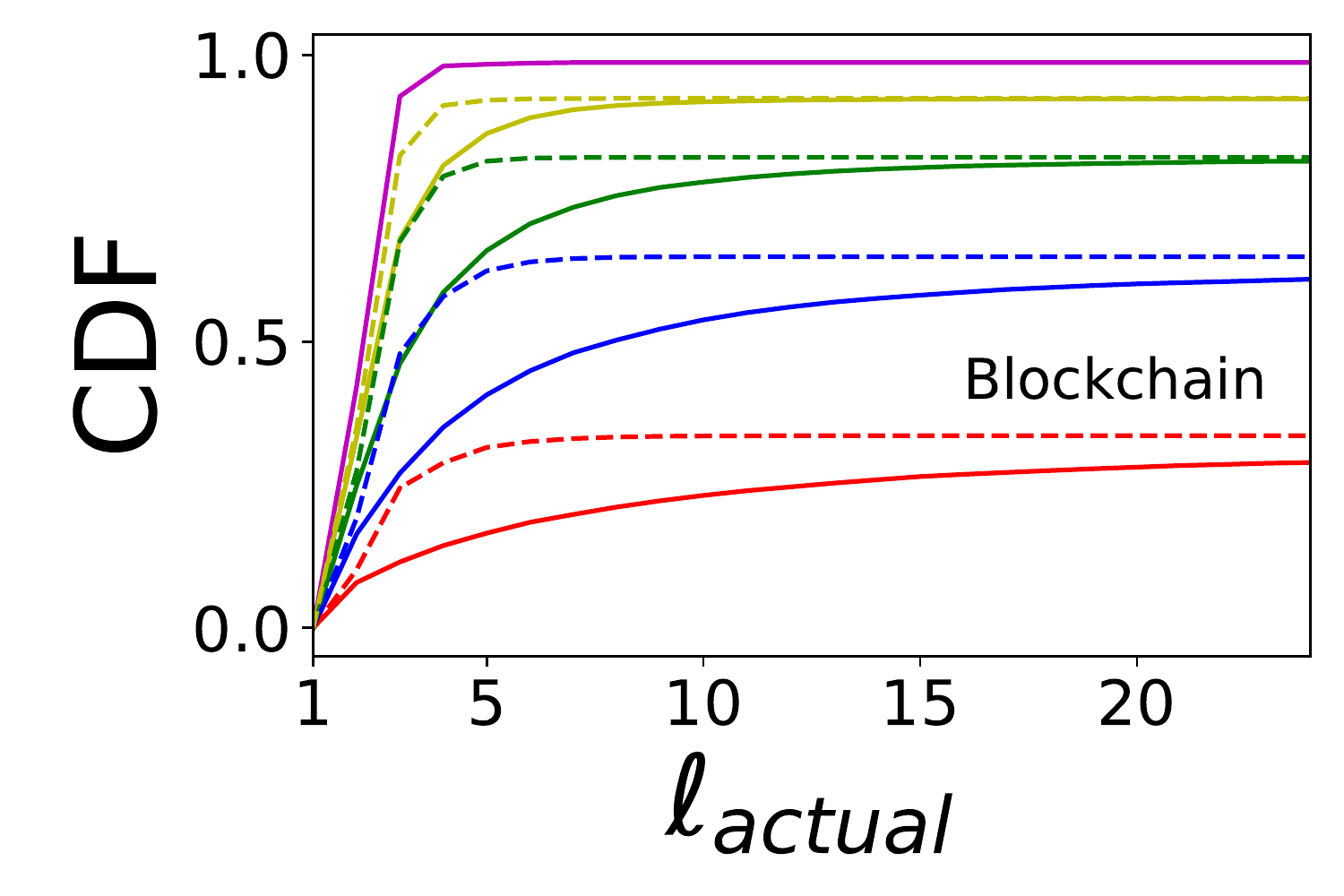}	
		\label{fig:act-cum-blockchain}
	}
	\hfill
	\subfloat{
		\includegraphics[width=0.31\textwidth]{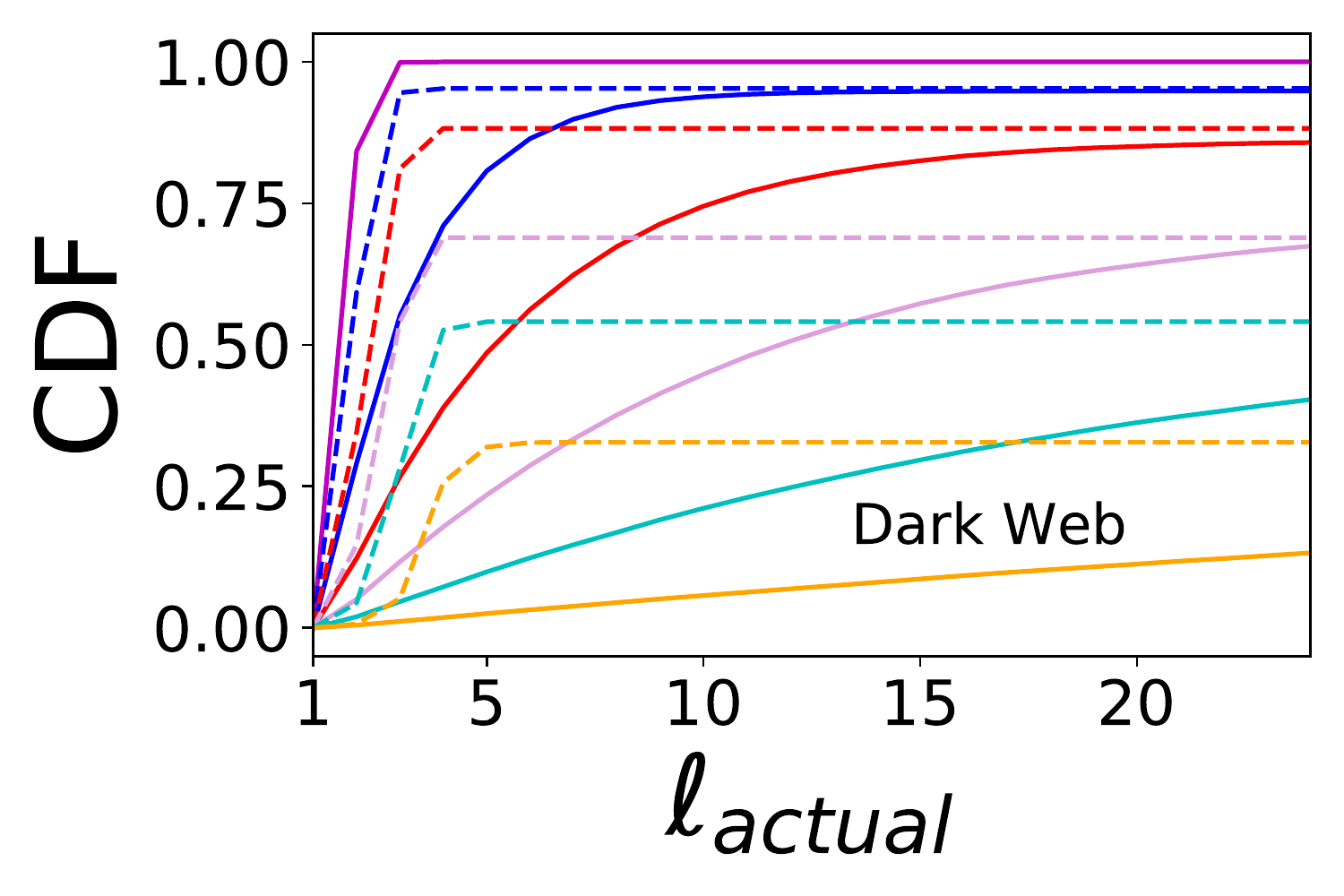}	
		\label{fig:act-cum-dacunha}
	}	
	\hfill	
	\subfloat{
		\includegraphics[width=0.31\textwidth]{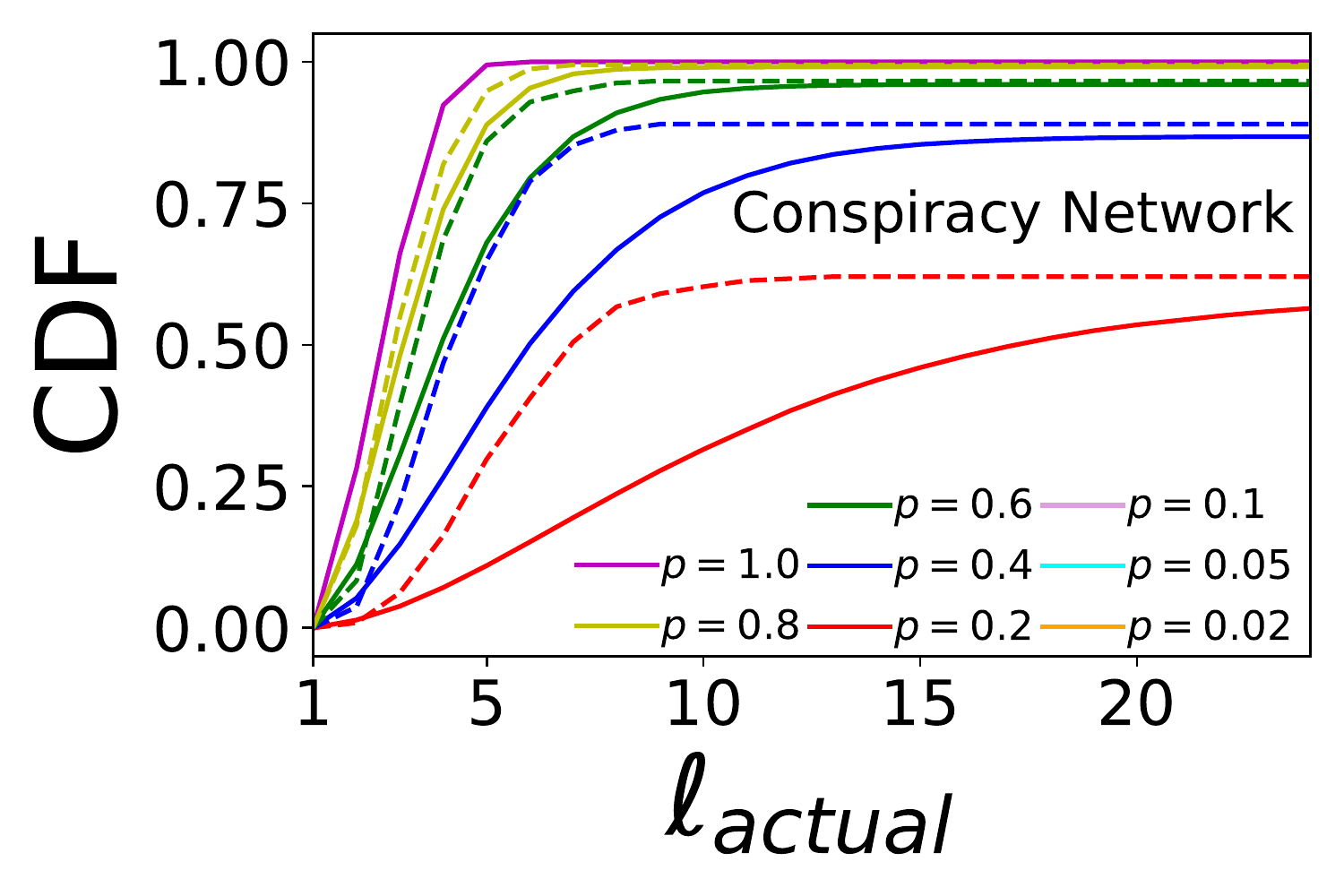}
		\label{fig:act-cum-matjaz}
	}

	\caption{\textbf{Targeting a specific node results.} We consider only nodes that are at least $\ell=2$ steps away and apply our greedy algorithm (Alg.~\ref{alg:greedy}) to deanonymize them given knowledge of a particular source node. Filled lines represent the results of our algorithm, whereas dashed lines represent the optimal number of nodes reached by an investigator with complete knowledge of the underlying network. We show the results for different values of $p$ in each network, with values of $p\leq0.1$ shown only for the Dark Web. We see that for all 3 of our networks, for high levels of $p$ our algorithm is very close to the optimal path (since most links involve exchange of identifiable information), whereas for lower values of $p$ far more nodes can be reached using optimal choices than our algorithm attains.  For these low levels of $p$, incorporating additional metadata could improve investigators' ability to deanonymize particular nodes. Results are from at least 20,000 simulated source and target pairs (for each value of $p$).   }
	\label{fig:act-cum}
\end{figure*}

We next apply our greedy algorithm to our three datasets. As before, for the Blockchain Transaction Network, we consider all of the exchange nodes as our source nodes (with all of their links known), whereas for the Dark Web Network and Conspiracy Network we choose source nodes randomly (and do not assume them to necessarily have any prior knowledge of all neighbors). We then choose 100 random target nodes and assess how many actual steps are required to reach them for different values of $p$. In order to understand the network effects, we focus on identifying nodes that are at least $\ell=2$ steps away from a source node.

In Fig.~\ref{fig:act-cum} we show the CDF of how many nodes are reached within $\ell_{actual}$ steps using our greedy algorithm and compare it to the number of nodes reached if the actual shortest path, $\ell$ were known. We find that the investigators' lack of knowledge about which links involve exchange of identifiable information is a significant detriment to their ability to optimally traverse the network for low values of $p$. For higher levels of $p$, the detriment is less pronounced as most of the links will actually involve identifiable information and thus attempting to traverse the shortest path will often succeed.

\begin{figure*}
	\centering

	\subfloat{
		\includegraphics[width=0.32\textwidth]{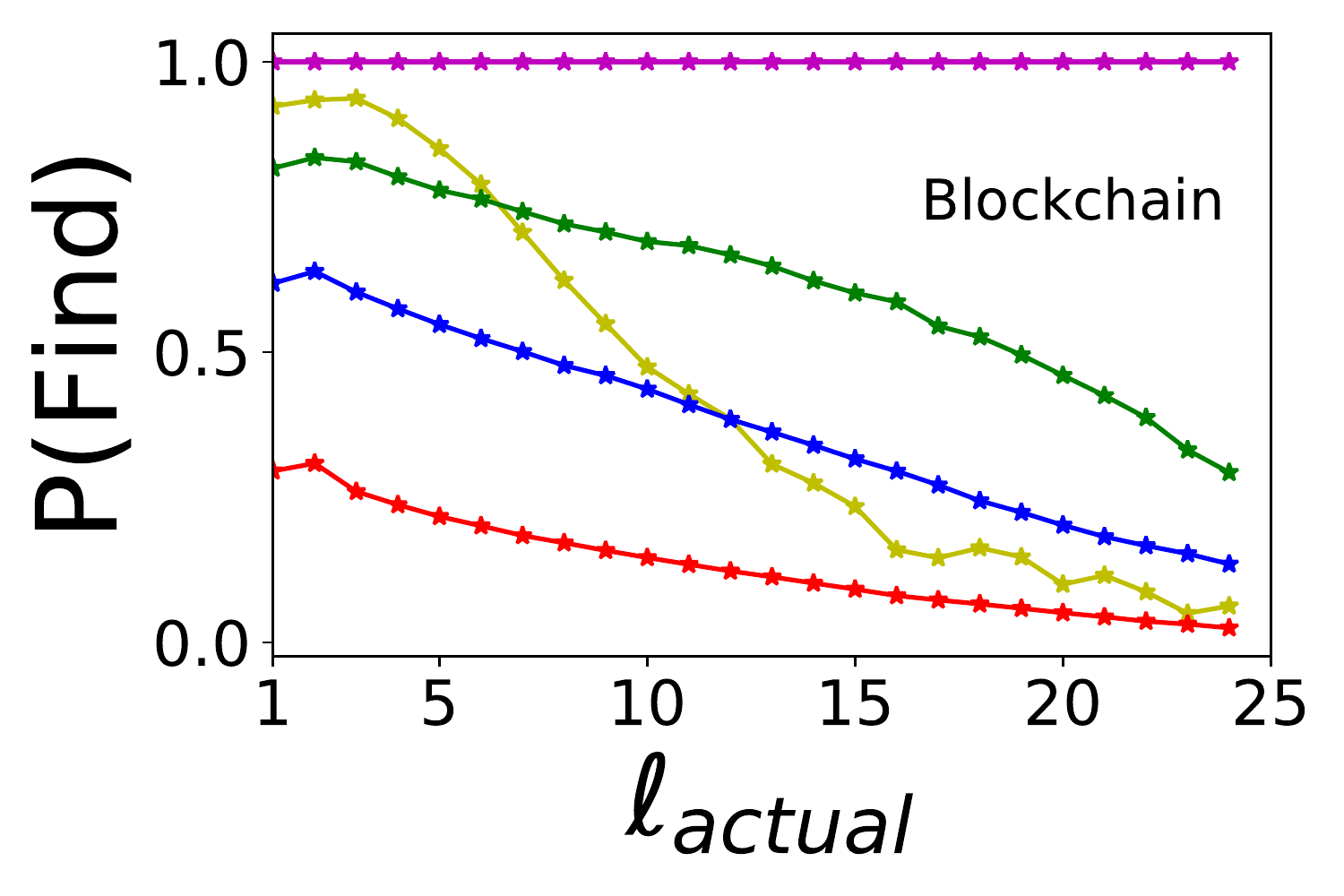}	
		\label{fig:will-find-blockchain}
	}
	\hfill
	\subfloat{
		\includegraphics[width=0.31\textwidth]{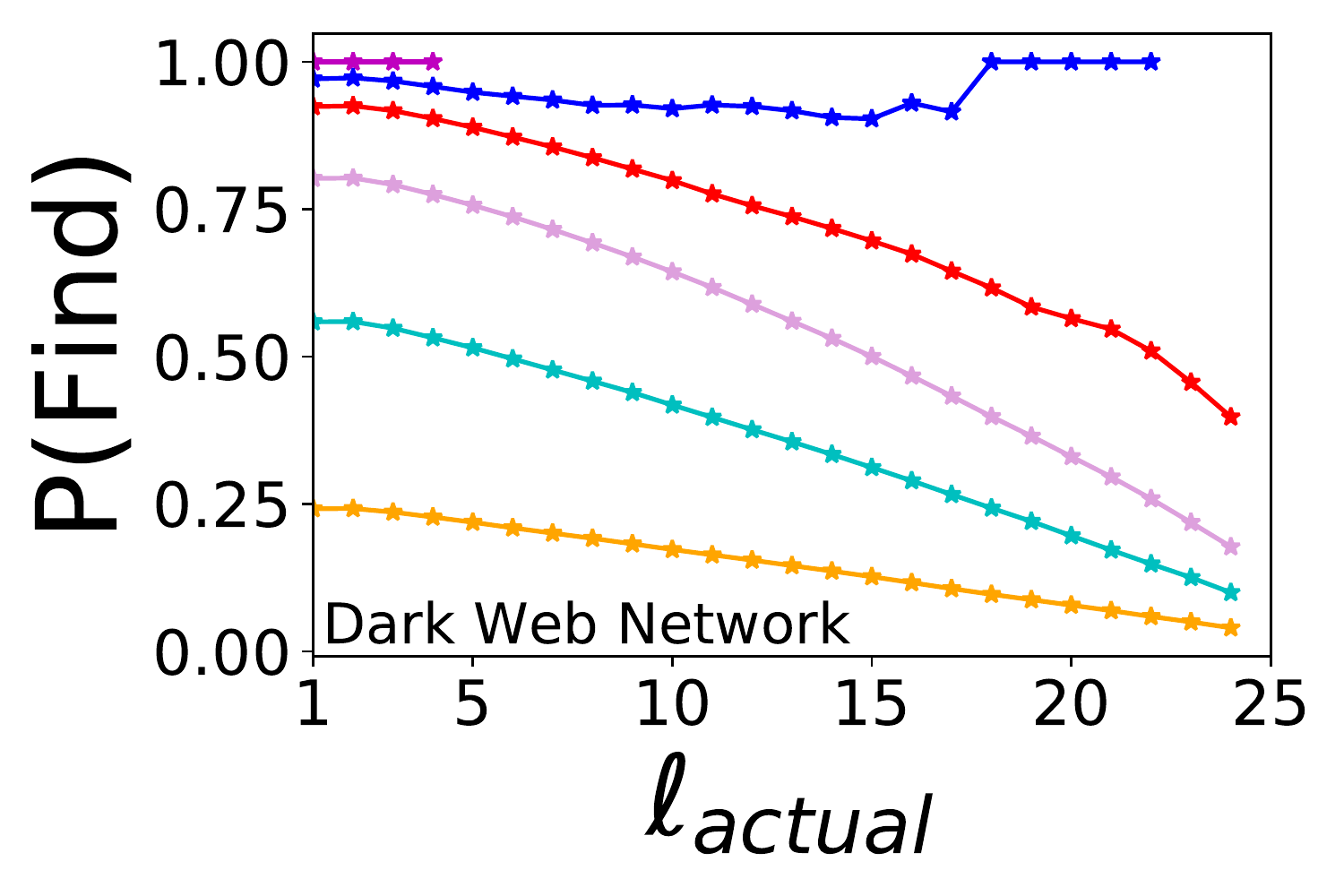}	
		\label{fig:will-find-dacunha}
	}	
	\hfill	
	\subfloat{
		\includegraphics[width=0.32\textwidth]{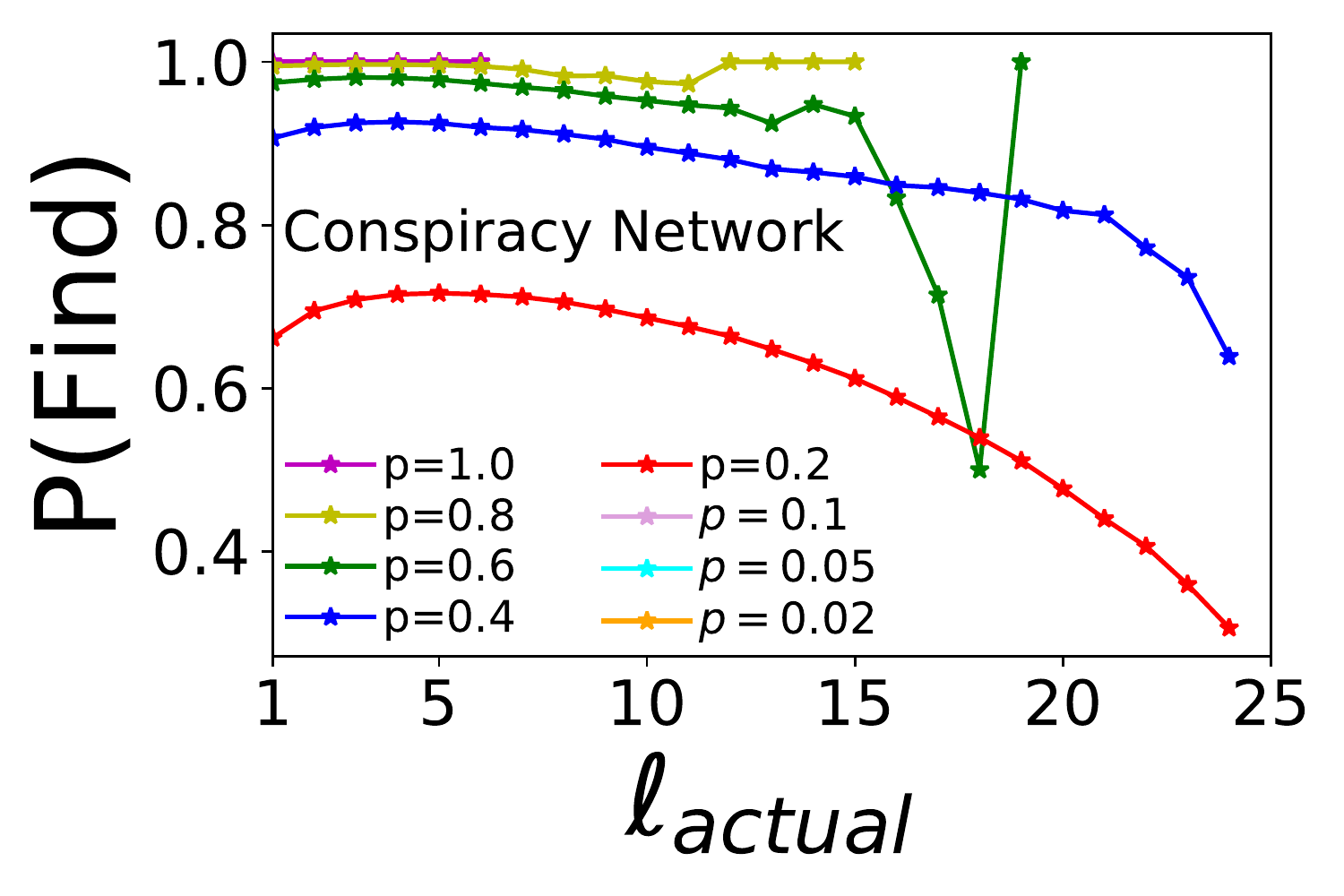}
		\label{fig:will-find-matjaz}
	}

	\caption{\textbf{Likelihood of future success.} Here we show for all three of our networks the likelihood that continuing to search after $\ell_{actual}$ steps will lead to a successful identification of a given target node. In particular for the Blockchain network, we see that for $p=0.8$ there is sharp decrease in the value of continued searching part around $\ell_{actual}\approx 6$ steps. For the other networks, we observe high probabilities of success except for $p=0.2$ where the chance of finding a target note, P(find), decreases steadily. Note that for larger values of $\ell_{actual}$ there are very small sample sizes leading to higher fluctuations. Likewise, for some values of $p$ the lines end before $\ell_{actual}=25$, because all nodes are either found or known to be unreachable before 25 steps. }
	\label{fig:will-find}
\end{figure*}

This detriment is further observed when we consider an investigators likelihood of success from continuing to investigate additional individuals. In other words, given that an investigator has investigated $k$ individuals, what is the likelihood that continuing to investigate the $k+1$st individual (and so on if necessary) will lead to identifying the desired target? In Fig.~\ref{fig:will-find} we show this likelihood for differing levels of $p$. We see that for the Blockchain network, until  $\approx5$ investigations, the likelihood of continuing to investigate is worthwhile, while after that, the probability of ultimate success decreases. Also, it is worth noting that for $p=0.8$, the success rate actually drops faster than for lower $p$. This is because for higher $p$, more individuals will be identified thus leading investigators to believe that they still may succeed, while for lower $p$ they will more quickly determine that the target is not identifiable (that is, the target is outside the giant component). For the other networks, we see that the value in continuing to investigate depends on $p$, as for high $p$, the likelihood of success continues to be significant, while for lower $p$ it drops fairly quickly.

We  note that our greedy algorithm does not use any metadata that may be associated with the links, which could improve an investigators ability to guess which links involved exchange of identifiable information. For example, one could look at the frequency that the different links appear e.g. how often transactions are made or how often two individuals communicate, and thus improve the success rate. Incorporating such information into a scoring system for the links could lead to improved values of $\ell_{actual}$.

\section{Discussion} 
We have demonstrated the fundamental mapping between percolation and privacy. Our framework can be used to understand networks where the actors (nodes) interact anonymously. Nonetheless, within the links (or at least within some of them) there often exists exchange of information that could allow for deanonmyization. This naturally leads to our percolation mapping with corresponding interpretations for the size of the giant component, number of components, and shortest path lengths.
We observed an existence of hubs in all three of our networks, which is a topological structure that is common in many networks. Importantly, we note that the existence of hubs exacerbates privacy issues as hubs can be potentially identified via their many spokes and then can reveal information on the remainder of the network. This is particularly true in the case of the Blockchain Transaction network where the hubs are exchanges which are known to collect information on their neighbors.


We further developed a greedy algorithm to account for the fact that the shortest path is not easily discovered and pursuing what appears to be the shortest path can often lead to a dead end. Thus, our greedy algorithm demonstrates the feasibility of actually effort that is required for discovering the identity of a chosen target node. Overall, our results provide a framework for investigators to approach deanonymization and consider their likelihood of success and required effort in order to deanonymize a particular node in the network or a given desired fraction of the entire hidden network. 

\section{Acknowledgement} 
This work was supported by the BIU Center for Research
in Applied Cryptography and Cyber Security in conjunction with the Israel National Cyber Directorate in the Prime Minister’s Office and also the Ariel Cyber Innovation Center in conjunction with the Israel National Cyber directorate in the Prime Minister's Office.



\bibliographystyle{unsrtnat}
\bibliography{ethereum}

\begin{thebibliography}{25}
\providecommand{\natexlab}[1]{#1}
\providecommand{\url}[1]{\texttt{#1}}
\expandafter\ifx\csname urlstyle\endcsname\relax
  \providecommand{\doi}[1]{doi: #1}\else
  \providecommand{\doi}{doi: \begingroup \urlstyle{rm}\Url}\fi

\bibitem[Gross and Acquisti(2005)]{gross2005information}
Ralph Gross and Alessandro Acquisti.
\newblock Information revelation and privacy in online social networks.
\newblock In \emph{Proceedings of the 2005 ACM workshop on Privacy in the
  electronic society}, pages 71--80, 2005.

\bibitem[Garcia(2017)]{garcia2017leaking}
David Garcia.
\newblock Leaking privacy and shadow profiles in online social networks.
\newblock \emph{Science advances}, 3\penalty0 (8):\penalty0 e1701172, 2017.

\bibitem[Bagrow et~al.(2019)Bagrow, Liu, and Mitchell]{bagrow2019information}
James~P Bagrow, Xipei Liu, and Lewis Mitchell.
\newblock Information flow reveals prediction limits in online social activity.
\newblock \emph{Nature human behaviour}, 3\penalty0 (2):\penalty0 122, 2019.

\bibitem[Reid and Harrigan(2013)]{reid2013analysis}
Fergal Reid and Martin Harrigan.
\newblock An analysis of anonymity in the bitcoin system.
\newblock In \emph{Security and privacy in social networks}, pages 197--223.
  Springer, 2013.

\bibitem[Lazer et~al.(2009)Lazer, Pentland, Adamic, Aral, Barab{\'a}si, Brewer,
  Christakis, Contractor, Fowler, Gutmann, et~al.]{lazer2009computational}
David Lazer, Alex Pentland, Lada Adamic, Sinan Aral, Albert-L{\'a}szl{\'o}
  Barab{\'a}si, Devon Brewer, Nicholas Christakis, Noshir Contractor, James
  Fowler, Myron Gutmann, et~al.
\newblock Computational social science.
\newblock \emph{Science}, 323\penalty0 (5915):\penalty0 721--723, 2009.

\bibitem[Song et~al.(2010)Song, Qu, Blumm, and Barab{\'a}si]{song2010limits}
Chaoming Song, Zehui Qu, Nicholas Blumm, and Albert-L{\'a}szl{\'o}
  Barab{\'a}si.
\newblock Limits of predictability in human mobility.
\newblock \emph{Science}, 327\penalty0 (5968):\penalty0 1018--1021, 2010.

\bibitem[Onnela et~al.(2007)Onnela, Saram{\"a}ki, Hyv{\"o}nen, Szab{\'o},
  Lazer, Kaski, Kert{\'e}sz, and Barab{\'a}si]{onnela2007structure}
J-P Onnela, Jari Saram{\"a}ki, Jorkki Hyv{\"o}nen, Gy{\"o}rgy Szab{\'o}, David
  Lazer, Kimmo Kaski, J{\'a}nos Kert{\'e}sz, and A-L Barab{\'a}si.
\newblock Structure and tie strengths in mobile communication networks.
\newblock \emph{Proceedings of the national academy of sciences}, 104\penalty0
  (18):\penalty0 7332--7336, 2007.

\bibitem[Zheleva and Getoor(2009)]{zheleva2009join}
Elena Zheleva and Lise Getoor.
\newblock To join or not to join: the illusion of privacy in social networks
  with mixed public and private user profiles.
\newblock In \emph{Proceedings of the 18th international conference on World
  wide web}, pages 531--540. ACM, 2009.

\bibitem[Xu and Chen(2004)]{xu2004fighting}
Jennifer~J Xu and Hsinchun Chen.
\newblock Fighting organized crimes: using shortest-path algorithms to identify
  associations in criminal networks.
\newblock \emph{Decision Support Systems}, 38\penalty0 (3):\penalty0 473--487,
  2004.

\bibitem[Barucca et~al.(2018)Barucca, Caldarelli, and
  Squartini]{barucca2018tackling}
Paolo Barucca, Guido Caldarelli, and Tiziano Squartini.
\newblock Tackling information asymmetry in networks: a new entropy-based
  ranking index.
\newblock \emph{Journal of Statistical Physics}, 173\penalty0 (3-4):\penalty0
  1028--1044, 2018.

\bibitem[Pappalardo et~al.(2018)Pappalardo, Di~Matteo, Caldarelli, and
  Aste]{pappalardo2018blockchain}
Giuseppe Pappalardo, Tiziana Di~Matteo, Guido Caldarelli, and Tomaso Aste.
\newblock Blockchain inefficiency in the bitcoin peers network.
\newblock \emph{EPJ Data Science}, 7\penalty0 (1):\penalty0 30, 2018.

\bibitem[Moody et~al.(2017)Moody, Galletta, and Dunn]{moody2017phish}
Gregory~D Moody, Dennis~F Galletta, and Brian~Kimball Dunn.
\newblock Which phish get caught? an exploratory study of individuals\'
  susceptibility to phishing.
\newblock \emph{European Journal of Information Systems}, 26\penalty0
  (6):\penalty0 564--584, 2017.

\bibitem[Newman(2010)]{newman2010networks}
Mark Newman.
\newblock \emph{Networks: an introduction}.
\newblock Oxford university press, 2010.

\bibitem[Cohen and Havlin(2010)]{cohen2010complex}
Reuven Cohen and Shlomo Havlin.
\newblock \emph{Complex networks: structure, robustness and function}.
\newblock Cambridge university press, 2010.

\bibitem[Castellano et~al.(2009)Castellano, Fortunato, and
  Loreto]{castellano2009statistical}
Claudio Castellano, Santo Fortunato, and Vittorio Loreto.
\newblock Statistical physics of social dynamics.
\newblock \emph{Reviews of modern physics}, 81\penalty0 (2):\penalty0 591,
  2009.

\bibitem[Barab{\'a}si et~al.(2016)]{barabasi2016network}
Albert-L{\'a}szl{\'o} Barab{\'a}si et~al.
\newblock \emph{Network science}.
\newblock Cambridge university press, 2016.

\bibitem[Chen et~al.(2018)Chen, Zhu, Li, Chen, Li, Luo, Lin, and
  Zhange]{chen2018understanding}
Ting Chen, Yuxiao Zhu, Zihao Li, Jiachi Chen, Xiaoqi Li, Xiapu Luo, Xiaodong
  Lin, and Xiaosong Zhange.
\newblock Understanding ethereum via graph analysis.
\newblock In \emph{IEEE INFOCOM 2018-IEEE Conference on Computer
  Communications}, pages 1484--1492. IEEE, 2018.

\bibitem[da~Cunha et~al.(2020)da~Cunha, MacCarron, Passold, dos Santos,
  Oliveira, and Gleeson]{da2020assessing}
Bruno~Requi{\~a}o da~Cunha, P{\'a}draig MacCarron, Jean~Fernando Passold,
  Luiz~Walmocyr dos Santos, Kleber~A Oliveira, and James~P Gleeson.
\newblock Assessing police topological efficiency in a major sting operation on
  the dark web.
\newblock \emph{Scientific Reports}, 10\penalty0 (1):\penalty0 1--10, 2020.

\bibitem[Ribeiro et~al.(2018)Ribeiro, Alves, Martins, Lenzi, and
  Perc]{ribeiro2018dynamical}
Haroldo~V Ribeiro, Luiz~GA Alves, Alvaro~F Martins, Ervin~K Lenzi, and
  Matja{\v{z}} Perc.
\newblock The dynamical structure of political corruption networks.
\newblock \emph{Journal of Complex Networks}, 6\penalty0 (6):\penalty0
  989--1003, 2018.

\bibitem[Ron and Shamir(2013)]{ron2013quantitative}
Dorit Ron and Adi Shamir.
\newblock Quantitative analysis of the full bitcoin transaction graph.
\newblock In \emph{International Conference on Financial Cryptography and Data
  Security}, pages 6--24. Springer, 2013.

\bibitem[Ober et~al.(2013)Ober, Katzenbeisser, and Hamacher]{ober2013structure}
Micha Ober, Stefan Katzenbeisser, and Kay Hamacher.
\newblock Structure and anonymity of the bitcoin transaction graph.
\newblock \emph{Future internet}, 5\penalty0 (2):\penalty0 237--250, 2013.

\bibitem[Meiklejohn et~al.(2013)Meiklejohn, Pomarole, Jordan, Levchenko, McCoy,
  Voelker, and Savage]{meiklejohn2013fistful}
Sarah Meiklejohn, Marjori Pomarole, Grant Jordan, Kirill Levchenko, Damon
  McCoy, Geoffrey~M Voelker, and Stefan Savage.
\newblock A fistful of bitcoins: characterizing payments among men with no
  names.
\newblock In \emph{Proceedings of the 2013 conference on Internet measurement
  conference}, pages 127--140. ACM, 2013.

\bibitem[De~Domenico and Arenas(2017)]{de2017modeling}
Manlio De~Domenico and Alex Arenas.
\newblock Modeling structure and resilience of the dark network.
\newblock \emph{Physical Review E}, 95\penalty0 (2):\penalty0 022313, 2017.

\bibitem[Sapovadia(2015)]{sapovadia2015legal}
Vrajlal Sapovadia.
\newblock Legal issues in cryptocurrency.
\newblock In \emph{Handbook of Digital Currency}, pages 253--266. Elsevier,
  2015.

\bibitem[Amaral et~al.(2000)Amaral, Scala, Barthelemy, and
  Stanley]{amaral2000classes}
Lu{\i}s A~Nunes Amaral, Antonio Scala, Marc Barthelemy, and H~Eugene Stanley.
\newblock Classes of small-world networks.
\newblock \emph{Proceedings of the national academy of sciences}, 97\penalty0
  (21):\penalty0 11149--11152, 2000.

\end{thebibliography}

\clearpage
\FloatBarrier
\setcounter{figure}{0}
\setcounter{table}{0}
\renewcommand{\thetable}{A.\arabic{table}}
\renewcommand{\thefigure}{A.\arabic{figure}}
\section{Supplementary Information}
\subsection{Data}
We collected our data from 3 public datasets corresponding to the cryptocurrency Ethereum \cite{chen2018understanding}, a network of interactions on the dark web \cite{da2020assessing}, and a political conspiracy network \cite{ribeiro2018dynamical}.

For the Ethereum network, we note that we specifically looked at what is referred to as the Money Flow Graph (MFG), which corresponds to the flow of funds through the network. Ethereum due to its design can also have nodes that are not related to the flow of money. We ignore these nodes and interactions between them since if no money is involved it is less likely that the ends of the two nodes will have identifiable information on one another. In the particular case of Ethereum in fact, many of these nodes are presumably `functions' that were written by a single developer but are perhaps used by many others.

We use the Dark Web network as provided by the authors \cite{da2020assessing}.

For the Conspiracy Network \cite{ribeiro2018dynamical}, we reduce the initial network to only include the GCC since the provided data involves actors in several conspiracy plots, some of which are not related to one another and thus represent distinct networks.

\begin{figure*}
	\centering

	\subfloat{
		\includegraphics[width=0.65\textwidth]{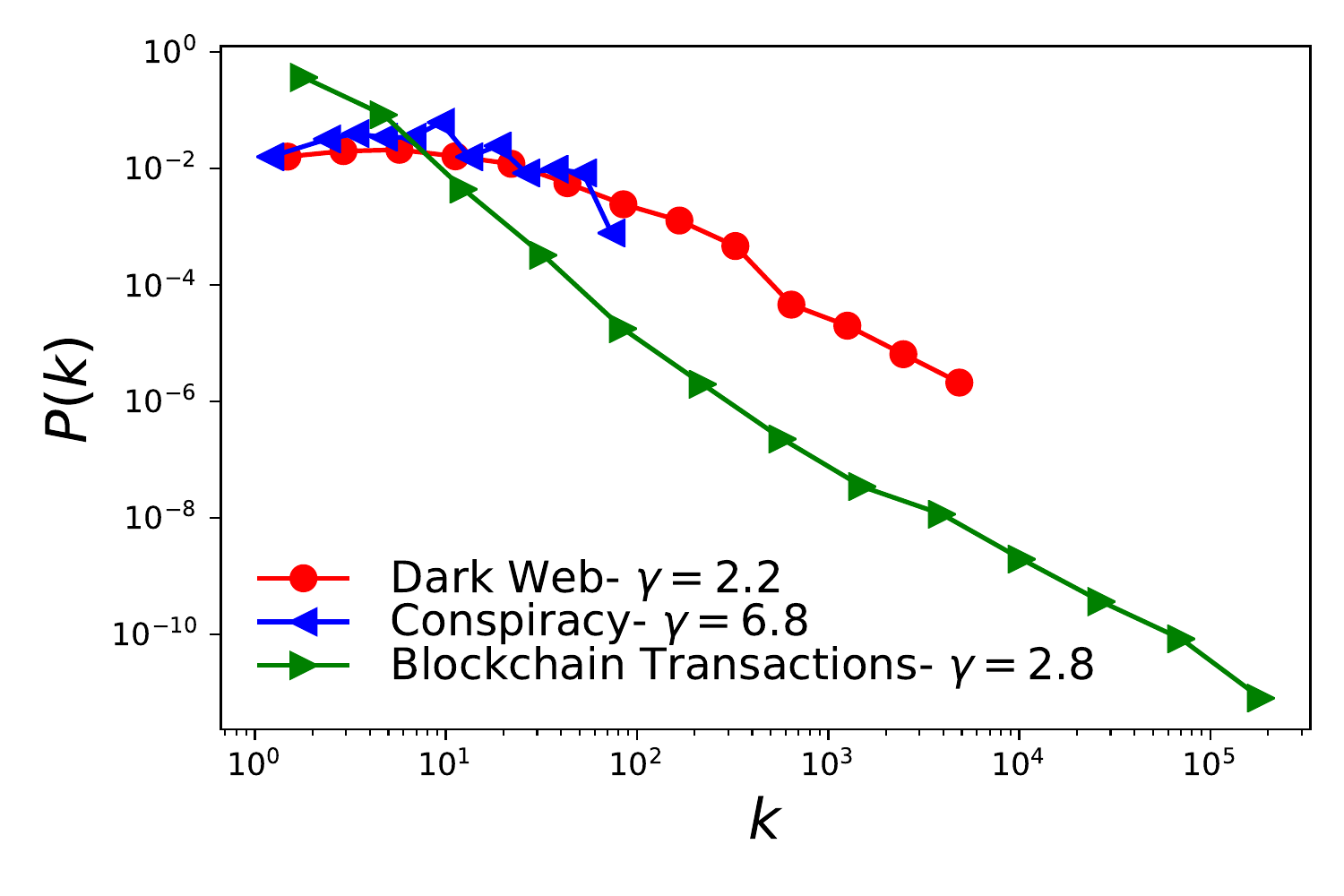}
		\label{fig:degree-dist1}
	}

	\caption{\textbf{Degree Distributions.} We show the degree distributions of our 3 networks. We see that all have degrees varying over several orders of magnitude. Recall that the Conspiracy and Dark Web networks are smaller and thus have a shorter tail. }
	\label{fig:degree-dist}
\end{figure*}

\subsection{Distribution of Component Sizes}
\begin{figure*}
	\centering
	\includegraphics[width=1.0\linewidth]{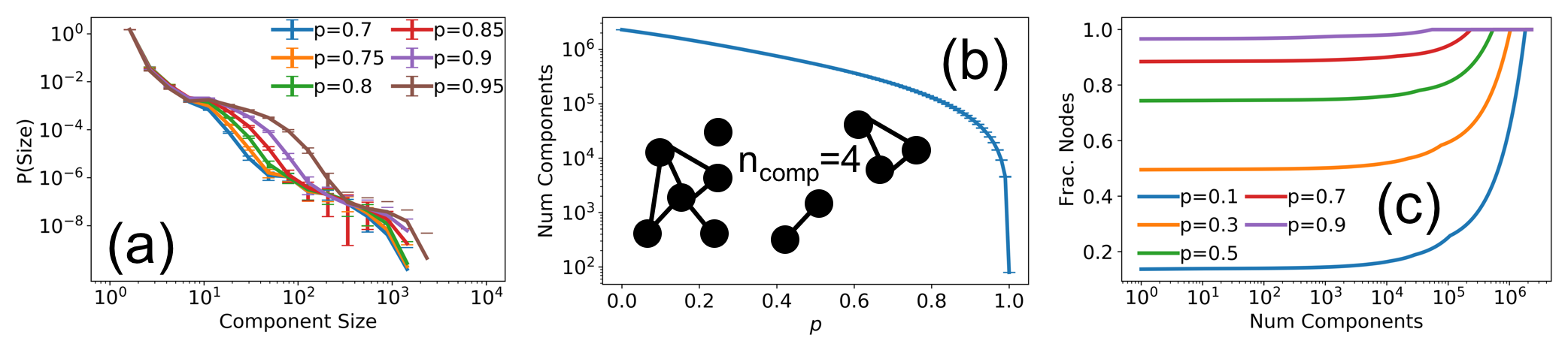}  	
	\caption{\textbf{Connected Components.} For deanonymization, an entire connected component is vulnerable if at least one if its nodes is deanonymized (as this node could then provide information on its neighbors and so on as in Fig.~\ref{fig:network-traverse}). Thus the number of components corresponds to the minimum number of deanonymized `seed' nodes in order for the entire network to be deanonymized. \textbf{(a)} Here we show the distribution of small components i.e., all components excluding the giant component as a function of $p$. We observe multiple scales of component sizes, however we do not observe power-law scaling for any values of $p$. \textbf{(b)} Here we show the total number of components (including isolated nodes as individual components) as a function of $p$. As $p$ decreases more nodes become isolated leading to a greater number of components. In the inset we show an example of 4 disconnected components, noting that the isolated node also counts as a component. \textbf{(c)} Here we show the total fraction of nodes reached after summing the $n$ largest components. We see that the total fraction of nodes reached starts at the size of the GCC and slowly increases. Note that the x-axis is log-scale thus, very large numbers of components are needed to reach beyond the GCC. }
	\label{fig:components}
\end{figure*}

An interesting point mentioned briefly in the main text, is that should a network consist of distinct components, an investigator would simply need a source node in each component in order to traverse the network and identify individuals. The Dark Web Network and Conspiracy Network are too small for us to observe a distribution of component sizes, however for the Blockchain Network, we show in Fig.~\ref{fig:components} several graphs related to component sizes. First we show the distribution of small components (Fig.~\ref{fig:components}a), which we see rapidly decreases with very few components above 10 nodes. We also show (Fig.~\ref{fig:components}b),  the number of components in the network for different values of $p$, which even for low values of $p$, may be many. However, many of these separate components are actually just a single isolated node, so their identification is difficult. To make this point clearer we show in Fig.~\ref{fig:components}c the fraction of nodes reached given that a varying number of components are identified. It is noteworthy that adding additional components has a minimal effect beyond the giant component until the number of components essentially covers all of the isolated nodes as well.

\subsection{Exchanges in the Blockchain Network}
 Our results in Fig.~\ref{fig:gcc-sgcc} suggested that all of our networks contains hubs. However, in the case of blockchain-based cryptocurrencies, these hubs are of a unique nature as prior work has noted that these hubs tend to be `exchanges' where one can convert cryptocurrency into ordinary fiat currencies \cite{chen2018understanding}. The exchanges have many more transactions than a typical consumer as they essentially serve as gatekeepers for their respective currencies. More importantly, nearly all of these exchanges now have know-your-customer (KYC) policies wherein they require passport information and other identifying information about their consumers \cite{sapovadia2015legal}. This means that transactions with an exchange inherently involve sharing identifiable information. 

To include the knowledge of exchanges in our percolation framework, we assumed that all exchange links involve identifiable information (due to KYC policies) and then assumed that $p$ fraction of the remaining transactions also involve sharing identifiable information. This framework was used throughout the main text when relating to the Blockchain Network. 

\begin{figure*}
	\centering
		\subfloat{
		\includegraphics[width=0.32\textwidth]{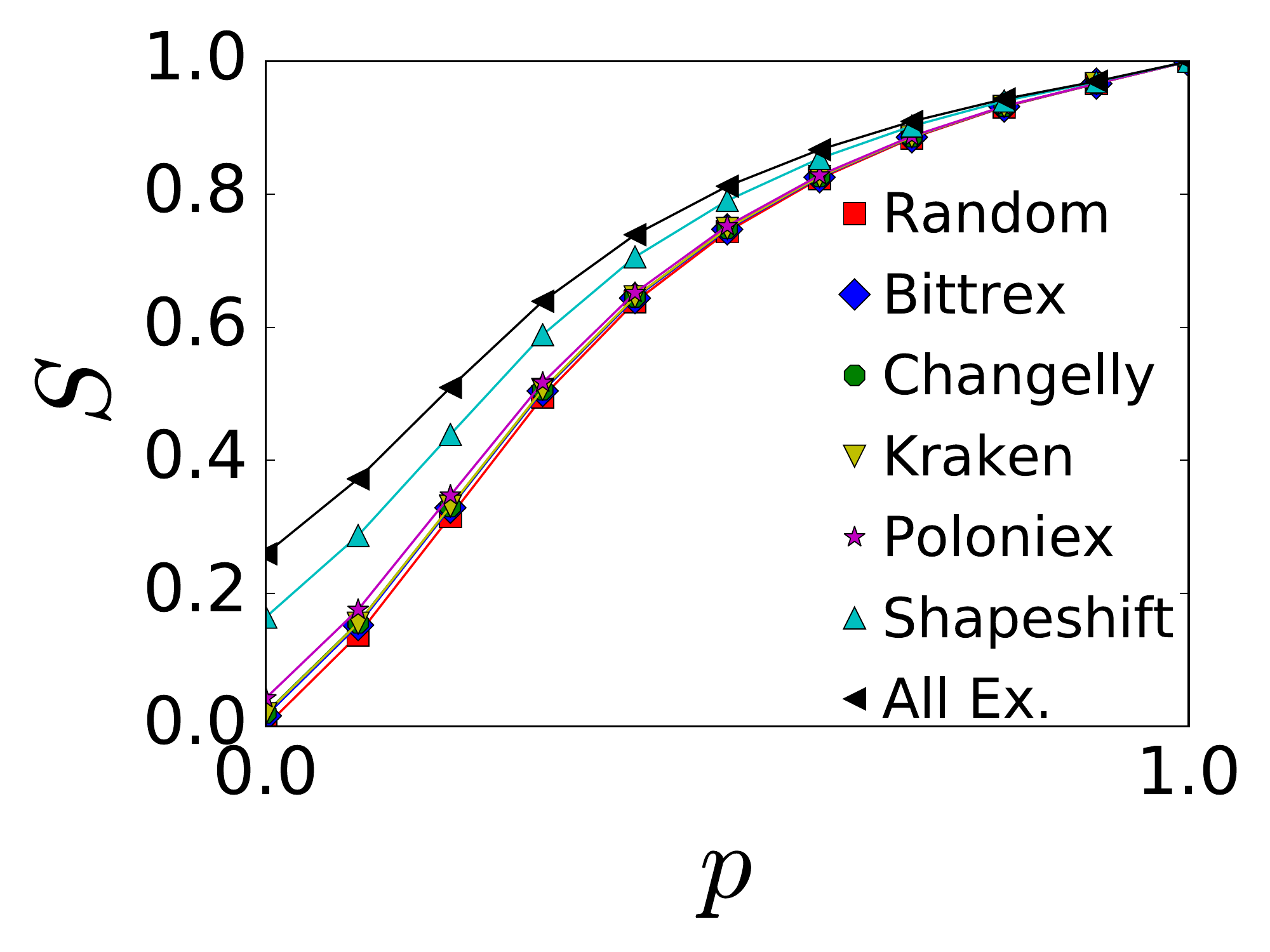}
		\label{fig:gcc-individual-exchange}
	}
	\hfill
	\subfloat{
		\includegraphics[width=0.32\textwidth]{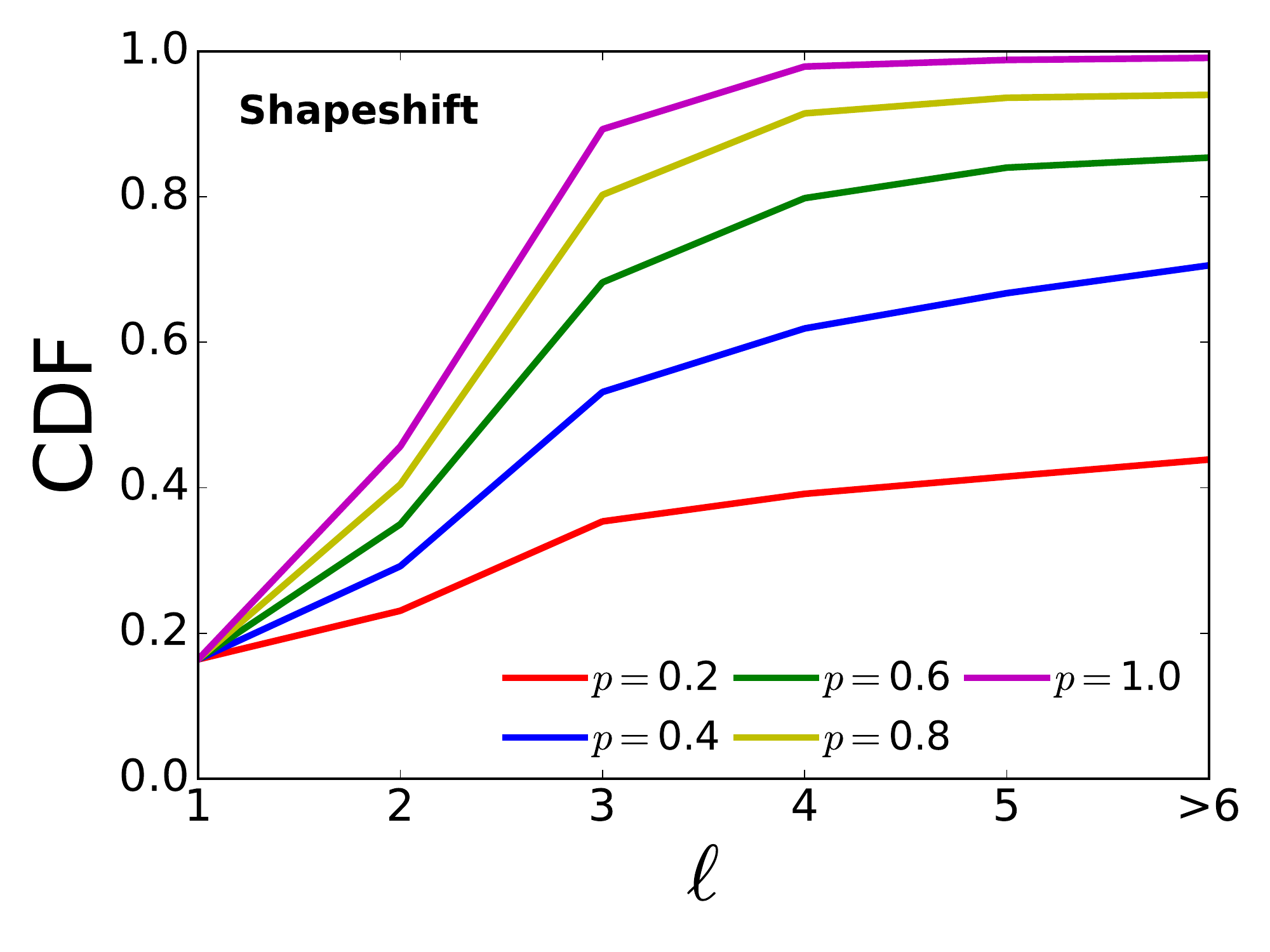}
		\label{fig:paths-shapeshift}
	}
	\hfill
	\subfloat{
		\includegraphics[width=0.31\textwidth]{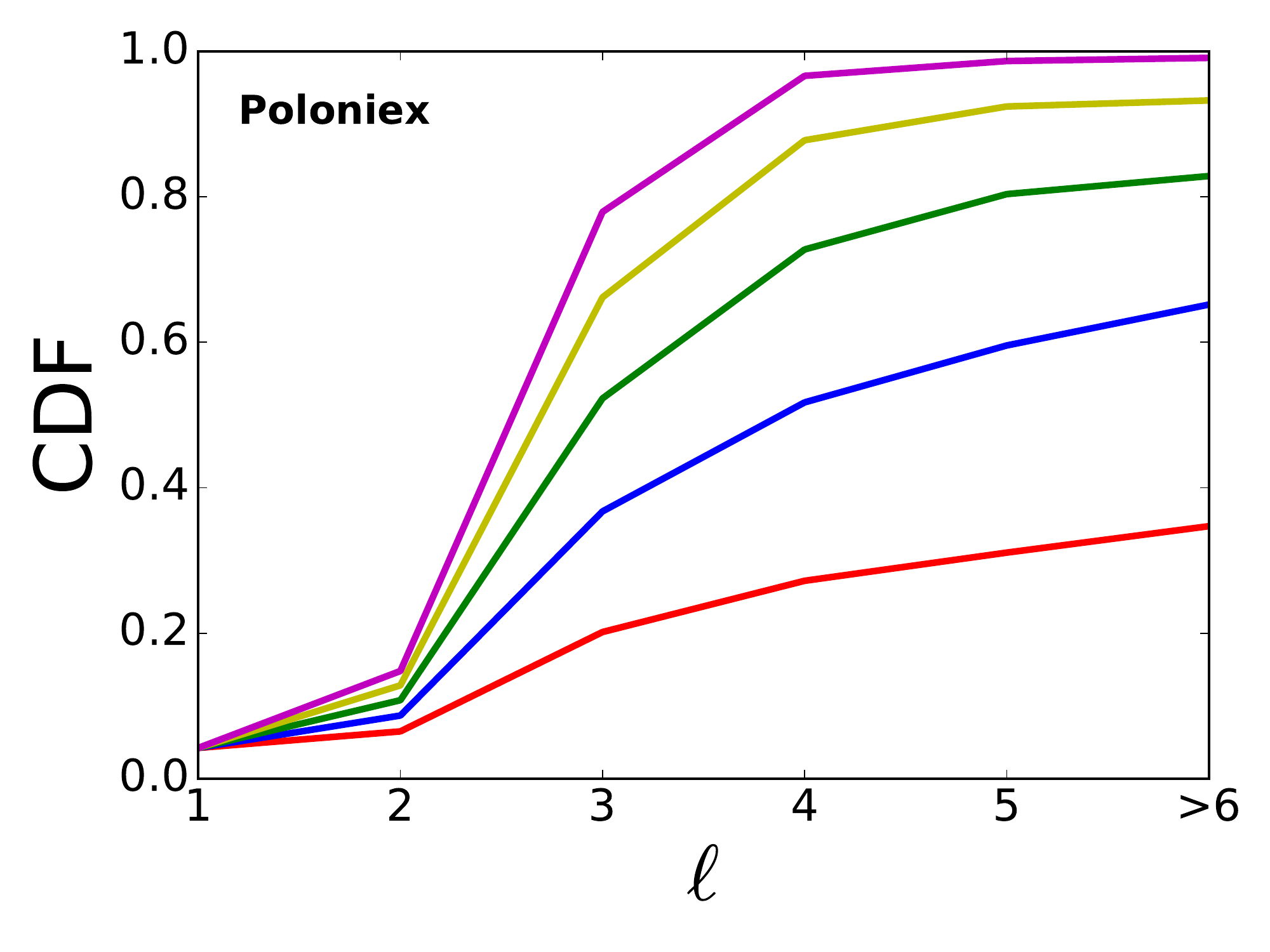}	
		\label{fig:paths-poloniex}
	}
	\vfill
	\subfloat{
		\includegraphics[width=0.32\textwidth]{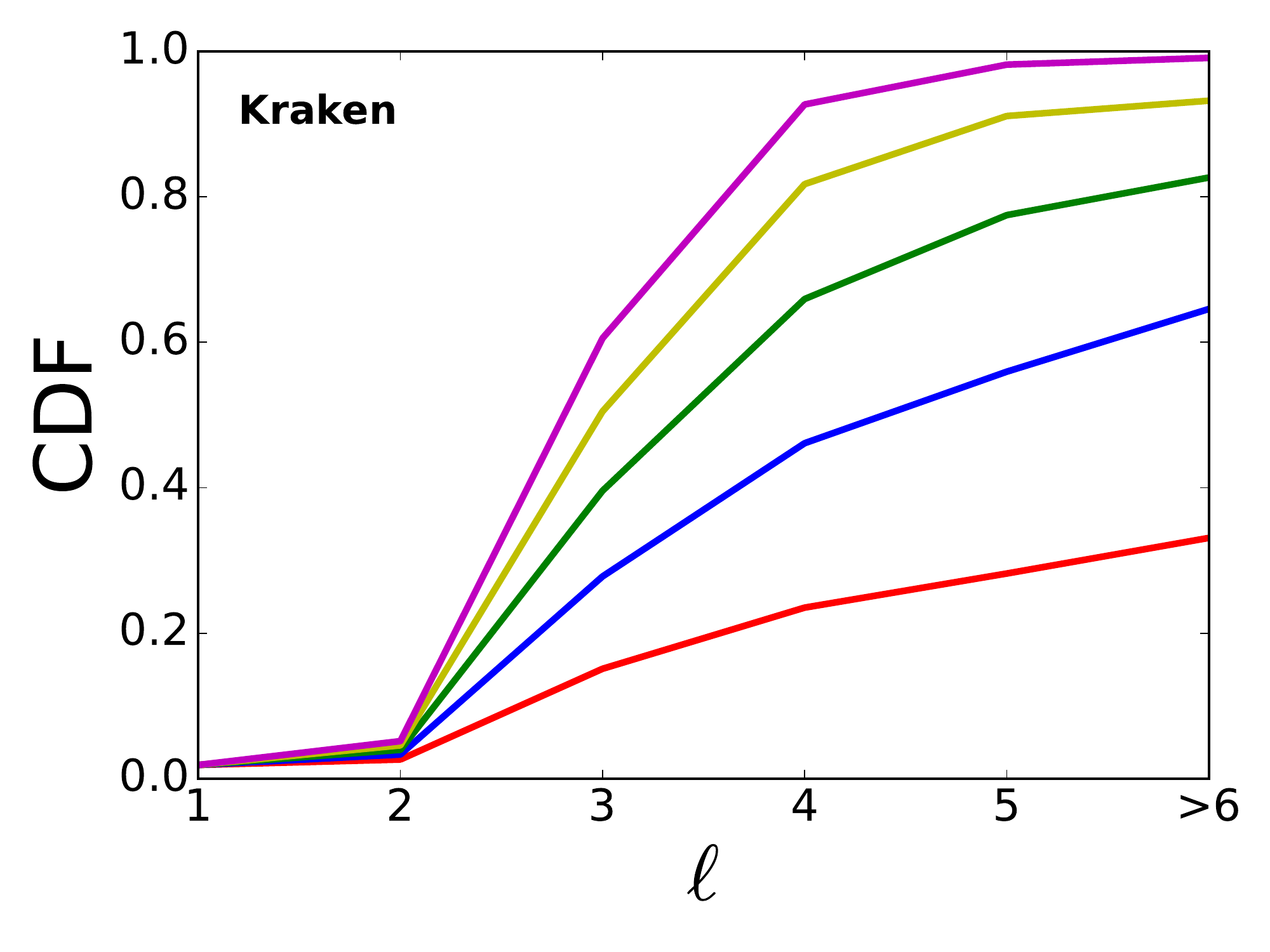} 
		\label{fig:paths-kraken}
	}\hfill
	\centering
	\subfloat{
		\includegraphics[width=0.32\textwidth]{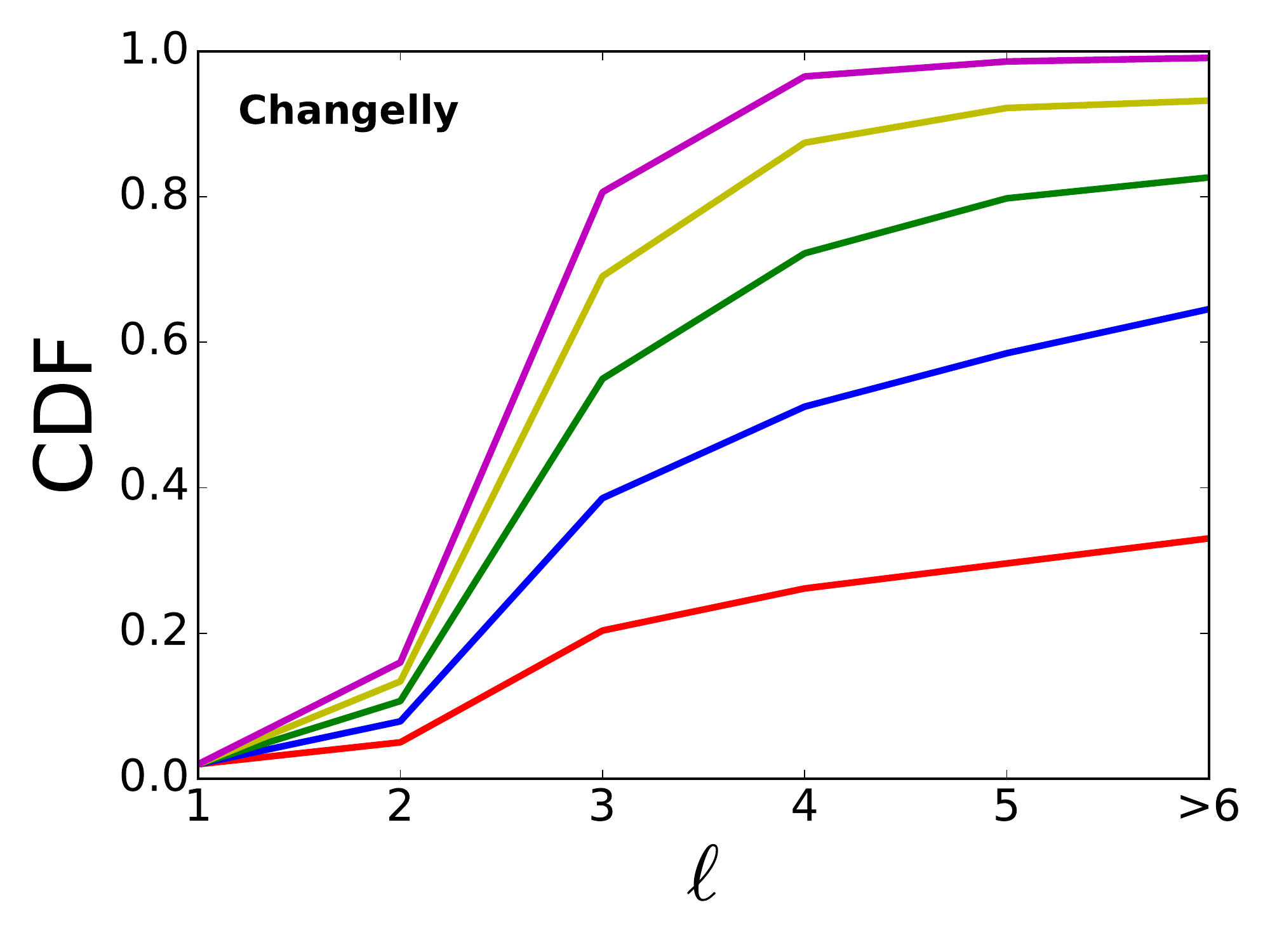}
		\label{fig:paths-changelly}
	}\hfill
	\subfloat{
		\includegraphics[width=0.31\textwidth]{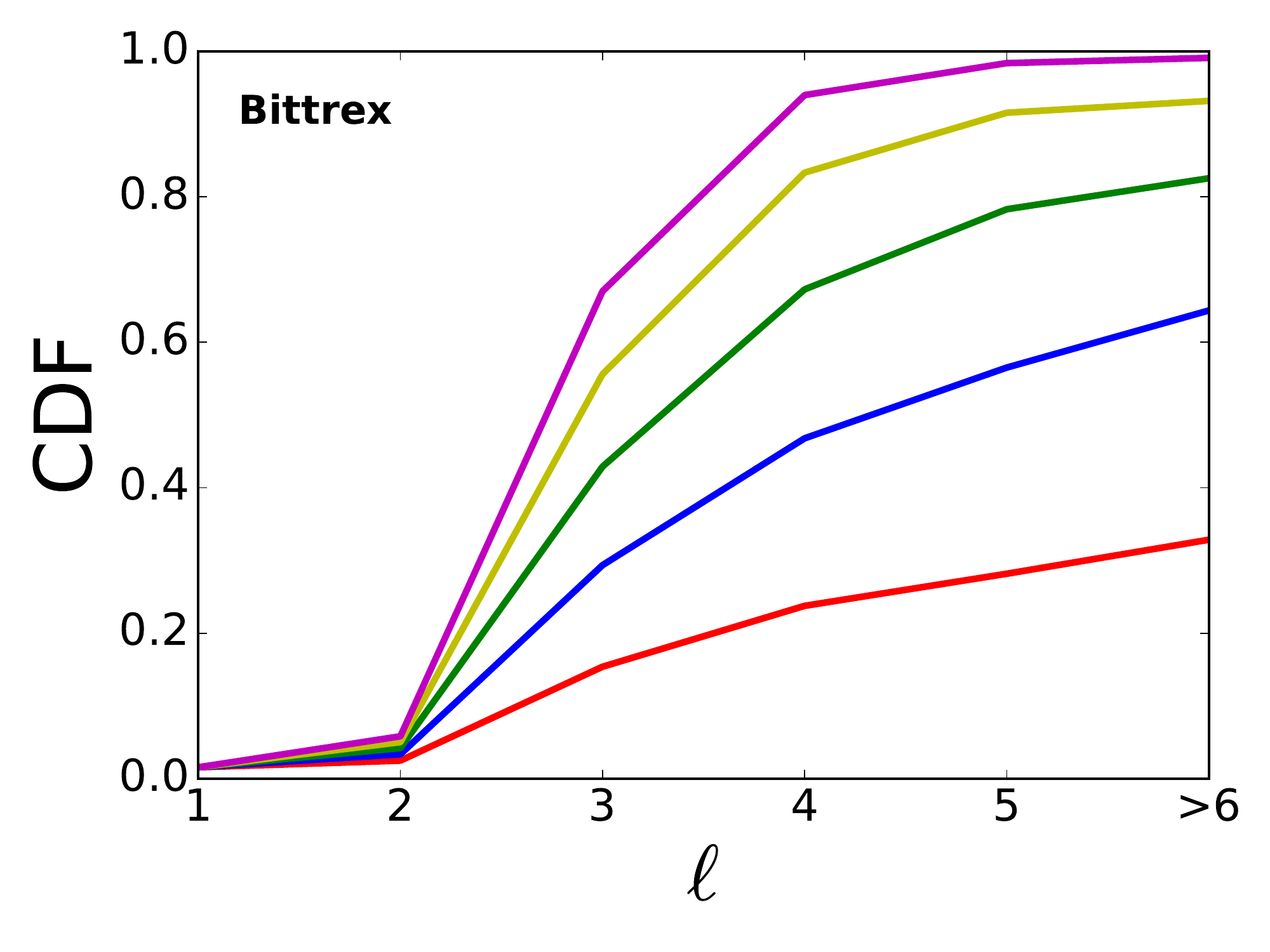}	
		\label{fig:paths-bittrex}
	}
	\caption{\textbf{Knowledge of links from only one individual exchange in the Blcockchain Network.} In the upper left we show the size of the giant component for the case where only links from an individual exchange (Bittrex, Changelly, Kraken, Poloniex, or Shapeshift are known) are guaranteed to be known. We see that knowing the links of Shapeshift has the largest effect for low values of $p$, but having knowledge from the other exchanges still leads to improvement as seen by the increase in $S$ for All exchanges. In the remaining five subplots we show the cumulative distribution (CDF) of nodes reached from an individual exchange node within $l$ steps. This is similar to Fig.~\ref{fig:spt-cdf} for the Blockchain Network, where all exchanges were assumed to be known, but here only the specific exchange in the label of the figure is known. We show for different values of $p$, what fraction of nodes are reached within $l$ steps. }
	\label{fig:paths-individual-exchanges}
\end{figure*}

Nonetheless, to expand on the results of the main text, we also include here the case where an investigator obtains information from only one of the exchanges i.e., Bittrex, Changelly, Kraken, Poloniex, or Shapeshift. We show both the sizes of the giant component and the CDF of the path lengths for each individual exchange in Fig.~\ref{fig:paths-individual-exchanges}.


%
%
%
%
%

We note that for the remaining networks (Dark Web and Conspiracy), we randomly selected a source node and did not assume that it necessarily knew the identities of its neighbors.

\FloatBarrier
 


\subsection{Actual Path lengths}
Here we present the probability distribution function (PDF) corresponding the CDF shown in Fig.~\ref{fig:act-cum}.
\FloatBarrier
\begin{figure*}
	\centering

	\subfloat{
		\includegraphics[width=0.32\textwidth]{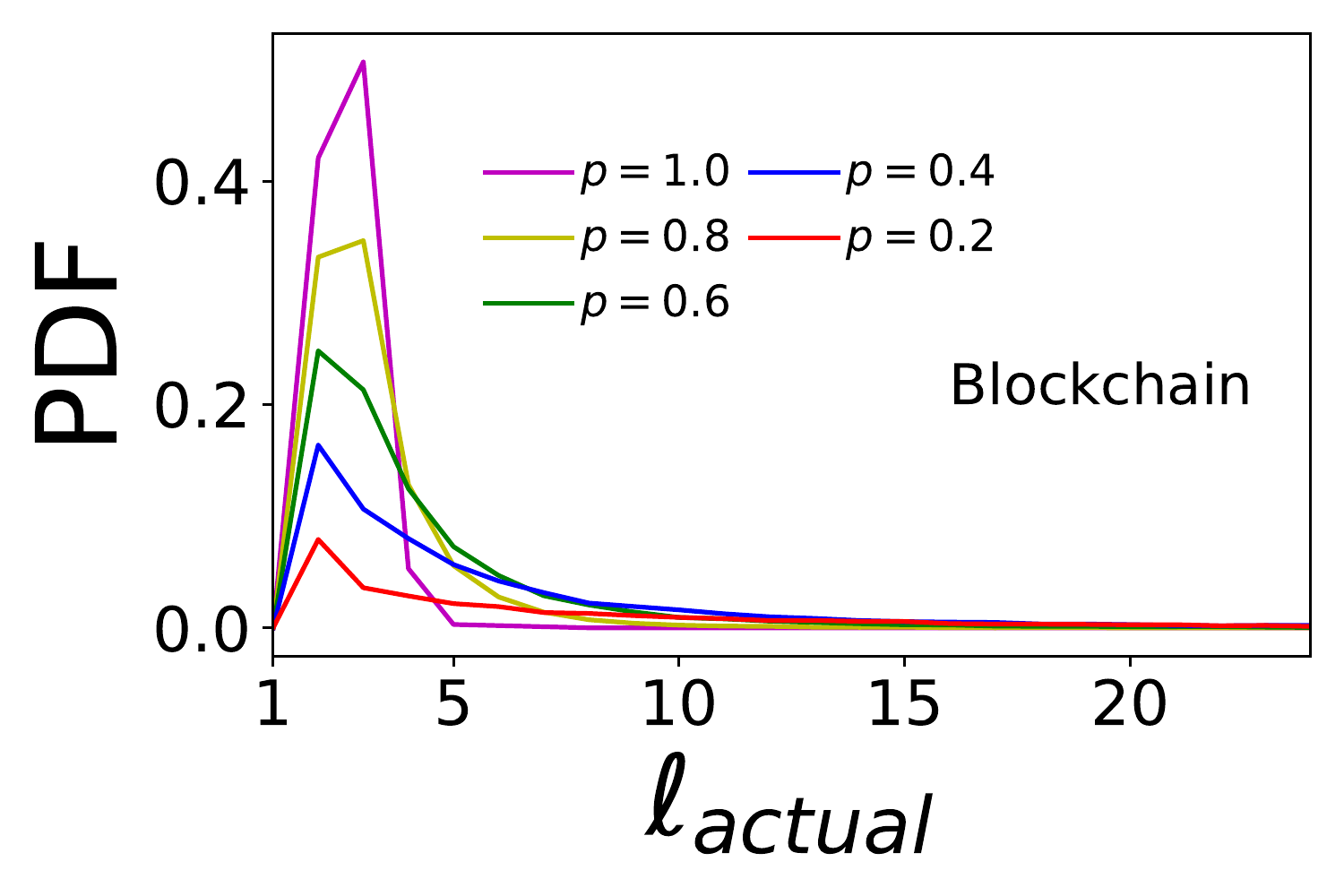}	
		\label{fig:actual-pdf-blockchain}
	}
	\hfill
	\subfloat{
		\includegraphics[width=0.31\textwidth]{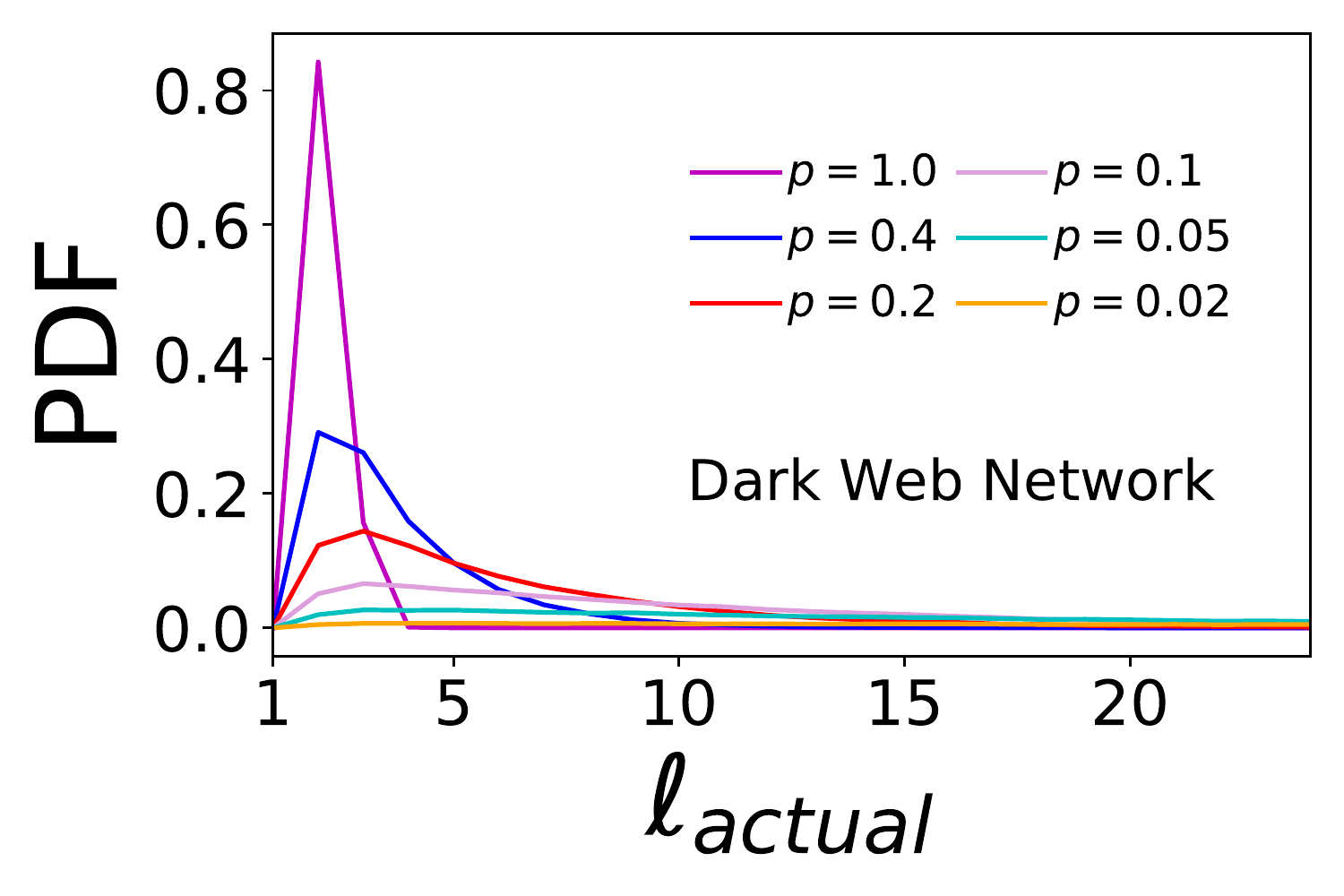}	
		\label{fig:actual-pdf-dacunha}
	}	
	\hfill	
	\subfloat{
		\includegraphics[width=0.32\textwidth]{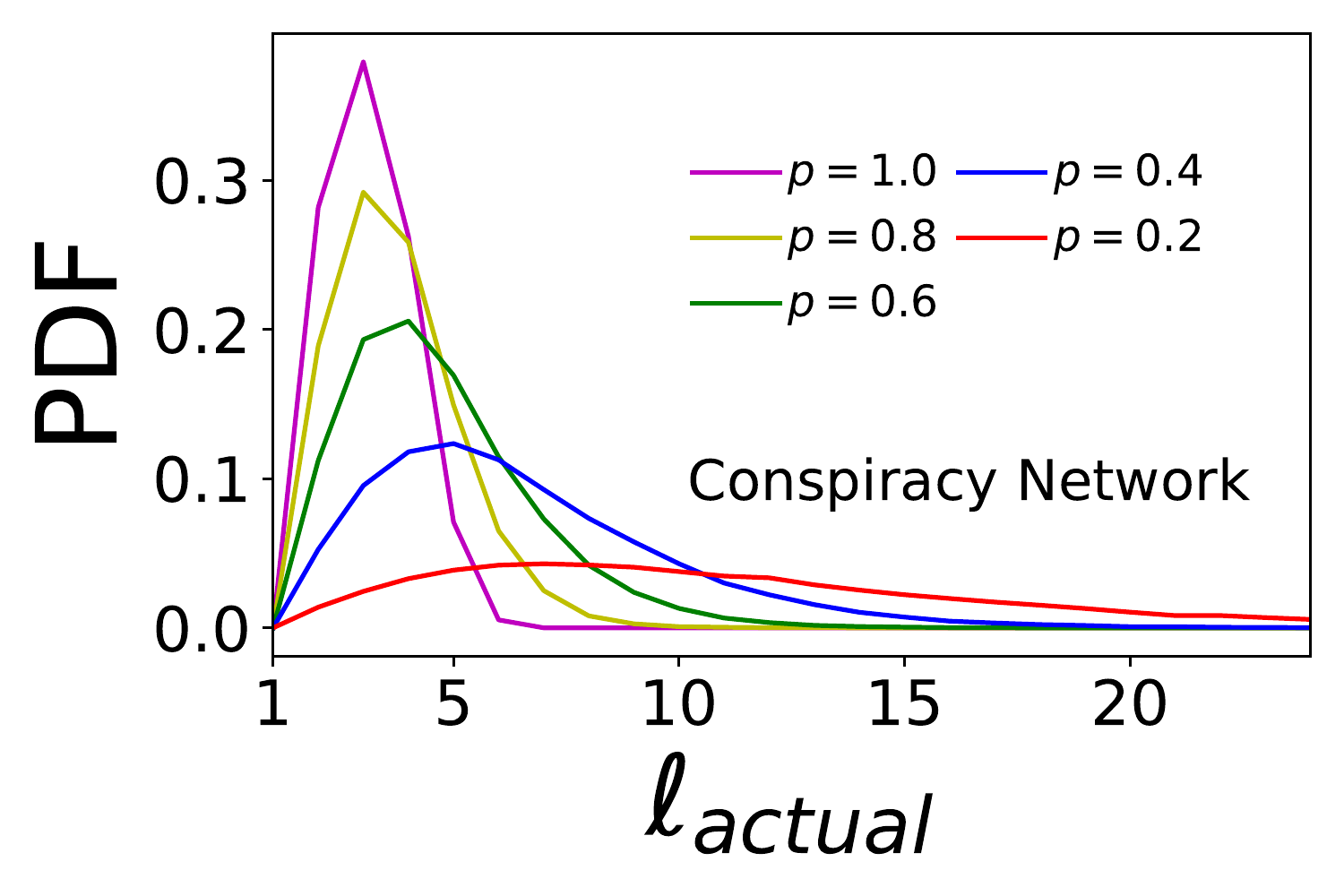}
		\label{fig:actual-pdf-matjaz}
	}

	\caption{\textbf{PDF of $\ell_{actual}$} Here we show the distribution of $\ell_{actual}$ that corresponds to the CDF shown in Fig.~\ref{fig:act-cum} of the main text. We observe a bell shaped curve with a peak around $\ell_{actual}\approx3$ for all of our networks. }
	\label{fig:actual-pdf}
\end{figure*}

\end{document}